\documentclass[fleqn,usenatbib]{mnras}
\usepackage{newtxtext,newtxmath}
\usepackage[T1]{fontenc}

\DeclareRobustCommand{\VAN}[3]{#2}
\let\VANthebibliography\thebibliography
\def\thebibliography{\DeclareRobustCommand{\VAN}[3]{##3}\VANthebibliography}

\usepackage{graphicx}	
\usepackage{amsmath}	
\usepackage{lipsum}
\usepackage{multirow}
\usepackage{float, graphicx, subfigure}
\usepackage{orcidlink}

\makeatletter
\DeclareRobustCommand{\HI}{%
  \mbox{H\check@mathfonts\fontsize\sf@size\z@\selectfont I}%
}

\renewcommand*{\@fnsymbol}[1]{\ensuremath{\ifcase#1\or ^{\ddag}\or \star\or \ast\or \dagger\or 
		\mathsection\or \mathparagraph\or \|\or **\or \dagger\dagger
		\or \ddagger\ddagger \else\@ctrerr\fi}}

\makeatother

\def\gN{\ensuremath{\bf g_{\scriptscriptstyle \rm N}}}
\def\gNabs{\ensuremath{g_{\scriptscriptstyle \rm N}}}
\def\PhiN{\ensuremath{\Phi_{\scriptscriptstyle \rm N}}}
\def\rhob{\ensuremath{\rho_\mathrm{b}}}
\def\phiph{\ensuremath{\phi_\mathrm{pdm}}}  
\def\rhoph{\ensuremath{\rho_\mathrm{pdm}}}

\def\grad{{\bf \nabla}}
\def\nut{{\tilde\nu}}

\def\msun{\ensuremath{M_{\odot}}}

\def\kms{\ensuremath{\mathrm{~km~s^{-1}}}}

\def\chisq{\ensuremath{\chi^2}}

\def\sigmastar{\ensuremath{\sigma_{\star}}}

\def\etal{\textit{et~al.}}

\def\QesComm#1{#1}  

\usepackage{ulem}

\def\kpc{\,\mathrm{kpc}}
\def\pc{\,\mathrm{pc}}

 \title[Testing gravitational models of galaxies]
 {How Close Dark Matter Halos and MOND Are to Each Other: Three-Dimensional Tests 
 	Based on Gaia DR2~\thanks{This work is evolved from two undergraduate-training projects, 
 		which were initially supported by Department of Astronomy of USTC and Yunnan Observatories of CAS, respectively.} }

\author[Y. Zhu, H.-X. Ma, X.-B. Dong et al.]{
{Yongda Zhu \orcidlink{0000-0003-3307-7525}}\,$^{1\dag}$,
Hai-Xia Ma \orcidlink{0000-0002-5237-9433}\,$^{2\dag}$,
Xiao-Bo Dong \orcidlink{0000-0002-2449-9550}\,$^{3}$\thanks{E-mail: xbdong@ynao.ac.cn}, 
Yang Huang \orcidlink{0000-0003-3250-2876}\,$^{4,5}$,
Tobias Mistele \orcidlink{0000-0001-7048-3173}\,$^{6}$, \newauthor~
Bo Peng\,$^{7,9}$, 
Qian Long\,$^{3}$, 
Tianqi Wang\,$^{8,9}$,
Liang Chang\,$^{3}$
and
Xi Jin $^{9}$
\\
$^{1}$Department of Physics \& Astronomy, University of California, Riverside, CA 92521, USA\\
$^{2}$Division of Particle and Astrophysical Science, Nagoya University, Nagoya, Aichi 464-8601, Japan\\
$^{3}$Yunnan Observatories, Chinese Academy of Sciences, Kunming, Yunnan 650011, China\\
$^{4}$University of Chinese Academy of Sciences, Beijing 100049, China \\  
$^{5}$National Astronomical Observatories, Chinese Academy of Sciences, Beijing 100101, China\\
$^{6}$Frankfurt Institute for Advanced Studies, Ruth-Moufang-Str. 1, D-60438 Frankfurt am Main, Germany\\
$^{7}$School of Information Engineering,
    Southwest University of Science and Technology,
    Mianyang, Sichuan 621010, China\\
$^{8}$Linx Lab, Hisilicon, Shenzhen, Guangdong 518129, China\\
$^{9}$Lab of Solid-State Electronics, Department of Physics,
University of Science and Technology of China, Hefei, Anhui 230026, China\\
$^{\dag}$ These authors contributed equally to this work.
}

\date{Accepted XXX. Received YYY; in original form ZZZ}

\pubyear{2022}

\begin{document}
\label{firstpage}
\pagerange{\pageref{firstpage}--\pageref{lastpage}}
\maketitle

\begin{abstract}
Aiming at discriminating different gravitational potential models of the Milky Way,
we perform tests based on the kinematic data powered by the Gaia DR2 astrometry,
over a large range of $(R,z)$ locations.
Invoking the complete form of Jeans equations that admit three integrals of motion,
we use the independent $R$- and $z$-directional equations  
as two discriminators ($T_R$ and $T_z$).
We apply the formula for spatial distributions of radial and vertical velocity dispersions
proposed by Binney \etal,  
and successfully extend it to azimuthal components, $\sigma_\theta(R,z)$ and $V_\theta(R,z)$;
the analytic form avoids the numerical artifacts caused by numerical differentiation in Jeans-equations calculation
given the limited spatial resolutions of observations, 
and more importantly reduces the impact of kinematic substructures in the Galactic disk.
It turns out that   
whereas the current kinematic data
are able to reject Moffat's Modified Gravity (let alone the Newtonian baryon-only model),
Milgrom's MOND is still not rejected.
In fact, both the carefully calibrated fiducial model invoking a spherical dark matter (DM) halo 
	and MOND are equally consistent with the data at almost all spatial locations
	(except that probably both have respective problems at low-$|z|$ locations),
no matter which a tracer population or which a meaningful density profile is used.
Because there is no free parameter at all in the quasi-linear MOND model we use, 
and the baryonic parameters 
are actually fine-tuned in the DM context,
such an effective equivalence is surprising, 
and might be calling forth a transcending synthesis of the two paradigms.

\end{abstract}




\section{Introduction}

Is the ``missing mass problem'' on (circum-)galactic scales 
due to the presence of dark matter (DM) or alternatively the delicate deviation of the underlying physical law from Newtonian gravity/dynamics? 
This is a fundamental and long outstanding question 
(see reviews, e.g., \citealt{milgrom_2010_review}, \citealt{feng_dark_2010}, \citealt{famaey_modified_2012},
\citealt{bullockSmallScaleChallengesLCDM2017a}  
and \citealt{Banik_Zhao_2022_review}).
The dynamics of gas and stars in and around galaxies has been observed to be 
in excess of the Newtonian gravity of the total baryonic content of the galaxies; 
the observational evidence includes the rotational curves of disk galaxies, the stellar velocity dispersion fields of low-luminosity galaxies and the low-surface-brightness parts of luminous galaxies, and so on (e.g.,  \citealt{angus_mass_2015,mcgaugh_radial_2016,dabringhausen_understanding_2016}).

Surprisingly and importantly, there are tight couplings 
(e.g., the Tully--Fisher relation; \citealt{Tully_Fisher_77_relation,McGaugh_2000_Baryonic_TF}) 
between the excess gravity and the baryonic content (see the above reviews). 
This fact inspired a gradually increasing number of researchers to interpret the ``excess'' with modified gravity (or dynamics) theories such as the ``modified Newtonian dynamics'' (MOND) proposed by \citet{milgrom_modification_1983} and the ``modified gravity'' (MOG) by \citet[][]{moffatScalarTensorVector2006}, instead of the popular DM paradigm. 
So far, however, no observational test is conclusive for the two competing paradigms on (circum-)galactic scales.

Previously, almost all observational tests \citep[see][]{famaey_modified_2012,Banik_Zhao_2022_review} of DM versus MOND (or MOG)
\footnote{\,By ``DM'' we mean Newtonian gravity with a DM component in addition to baryonic components for galaxies. 
	In the present paper we focus on the ``extra mass/gravity'' phenomena 
	on circum-galactic and galactic scales only, 
	i.e., within the gravitational binding of the so-called dark matter halo (in the DM language) hosting some galaxy. 
	It is in this context that the statements like this sentence hereinafter should be understood.}
have employed either rotational velocity data commonly (in disk galaxies) 
or sometimes stellar velocity dispersion (\sigmastar) data, with only a few exceptions 
(e.g., \citealt{angus_mass_2015}, \QesComm{\citealt{Lisanti_2019_PRD}, and \citealt{chrobakova_gaia-dr2_2020}}; 
see also \citealt{nipoti_vertical_2007} for methodological analysis) 
using both observed rotational curve (RC; in the galactic-disk plane) 
and observed \sigmastar\ information (particularly in the direction vertical to the disk) of a galaxy. 
By invoking Jeans equations, data of  \sigmastar\ as well as streaming velocity $\bar{v}_\theta$ are linked to 
models of the galactic gravitational potential $\Phi$ \citep[see \S4.8 of][]{binney_galactic_2008}. 
The advantage of jointly using both RC and \sigmastar\ data is obvious, with more constraints independently (\citealt{stubbs_testing_2005}).

Unfortunately, almost all the studies involving \sigmastar\ data in the literature 
adopted an unrealistic simplification of Jeans equations:
they all assumed a two-integral distribution function,
for instance, the popular $f = f(H, L_z)$ where $H$ is the Hamiltonian of the system and $L_z$ the $z$-direction angular momentum.
Thus, the stellar velocity-dispersion tensor having $\sigma_R = \sigma_z$ and $\sigma_{Rz} = 0$ 
denoted in the cylindrical coordinate system ($R, \theta, z$); 
i.e., the \sigmastar\ distribution in a meridional plane is isotropic, and the tilt angle of the velocity ellipsoid $\alpha = 0$. 
By doing so, the corresponding velocity-dispersion terms in Jeans equations are reduced or vanished accordingly, 
and the Jeans equations are closed (see \S2.1 of \citealt{nipoti_vertical_2007}, \S2.2 of \citealt{angus_mass_2015},
\QesComm{\mbox{\S\hspace{0.4pt}III.B} of \citealt{Lisanti_2019_PRD}}; cf. \S2.1 of \citealt{kipper_stellar_2016}). 
But, the fact, well known for decades, is that 
$\sigma_R \neq \sigma_z$ and $\sigma_{Rz} \neq 0$ in the observed disk galaxies 
(e.g., the MW and M31; see \citealt{kipper_stellar_2016} and references therein).
Besides, there is more evidence supporting the viewpoint that the stellar orbits do respect, 
for which there is no analytic expression though, a third integral of motion 
(see \citealt{kipper_stellar_2016}; also \S3.2 and \S4.4 of \citealt{binney_galactic_2008}).
\QesComm{Specifically, concerning Jeans-equations modeling of the MW, 
 the necessity of incorporating the cross-dispersion term $\sigma_{Rz}$ (i.e., tilt angle) in Jeans equations 
 has been thoroughly analyzed (e.g., \citealt{Hessman_2015_difficulty}, 
 \S3 of \citealt{Budenbender_2015_critic} and more subsequent studies).}

Besides the purpose to close Jeans equations, a practical reason of the above unrealistic simplification is 
to circumvent the calculating difficulty: the Jeans equations can only be solved \textit{numerically} 
for all practical purposes with observational kinematic data used, 
and---to be worse---it usually requires algorithmic techniques to calculate the general form of Jeans equations 
(involving three distinct \sigmastar\ components and the cross term $\sigma_{Rz}$ and their derivatives) 
given the \textit{limited} observational data so far.
Normally, it involves numerically calculating the partial derivatives of those \sigmastar\ components with respect to $R$ and $z$
(e.g., \citealt{chrobakova_gaia-dr2_2020}; cf. \S\ref{sec:our_kinematics_formulae}), 
which in principle demands dense sampling along the $R$ and $z$ directions 
(as well as careful numerical differentiation schemes or novel algorithms to minimize 
the notorious ``huge numerical errors''), 
and worse, is vulnerable to the impact of galactic substructures.  
The worrisome fact is that the stellar kinematics in galactic disks (e.g., the disk of the MW) is commonly affected by 
stellar substructures; or, in other words, 
galactic disks are full of kinematic substructures 
\citep{gaia_aa_2018_616_a11}.

In addition, in the aforementioned studies invoking Jeans equations, they not only simplified Jeans equations by assuming two-integral dynamics, 
but also usually approximated the solution of Jeans equations 
with an algebraic formula between the averaged vertical $\sigma^2_\star$ and the mass surface density {\it locally at every radius $R$}, 
i.e., $\sigma^2_z(R)$  $\sim \Sigma(R)$ 
(\QesComm{commonly seen for external disc galaxies}; see, e.g., \citealt{angus_mass_2015}). 
That simple formula was derived by neglecting the components of gravitational force in the $z=const$ planes 
(i.e., assuming that the gravitational force is in the $z$ direction only,  a so-called planar symmetry in the literature), 
which was actually wrongly assumed (or over-simplified) for the dynamics of stars 
\QesComm{
(see \S6.1 of \citealt{Piffl_Binney_etal_2014MN445} and \S4.2 of \citealt{McGaugh_2016_RCgradient} for the MW;
Footnote~2 of \citealt{nipoti_vertical_2007} for external galaxies).}
This simplification is actually the simplest version of the old ``$K_z$ method'' so-called in the literature, 
and makes the system completely one-dimensional, 
in the sense of both Jeans equations and Poisson equation.
\QesComm{To be specific, following the notations of \citet{Read_2014_JPhG_41} (see his \S3.3),
	while $K_z(z)$ means vertical force, \mbox{$ - \frac{\partial \Phi}{\partial z}$\/$(z)$} literally, 
this simplest $K_z$ method yet ignores both the ``tilt term'' in $z$-directional Jeans equation
and the ``rotation-curve term'' in Poisson equation.
Likewise, in some studies using MW data, the link between the vertical density profile ($\rho(z)$\,) 
and vertical distribution of the $z$-component velocity dispersion ($\sigma_z(z)$\,) 
of a tracer population was established by this simplest $K_z$ method 
(see, e.g., \mbox{\S\hspace{0.4pt}III.B} of \citealt{Lisanti_2019_PRD}). 
}

Aiming at observationally discriminating between DM and alternative gravitational potential models, 
we employ the complete form of Jeans equations that admit three integrals of motion,
\QesComm{
\footnote{Regarding the methodology, one of our aspirations came from the critical analysis 
	by \mbox{M.\,Milgrom} on the methodology of the DiskMass project, 
	particularly on the analysis method of \citet{angus_mass_2015}; 
	see \citet{Milgrom_arXiv_critical} and \citet{Angus_etal_2016_critical} for the detail.}
} 
and perform tests on the latest kinematic data powered by the Gaia DR2 astrometry. 
In the Gaia era, the measurement uncertainties (e.g., the effect onto kinematic quantities 
caused by systematic bias in distance estimation)
are no longer the major concern (see \S\ref{sec:our_kinematics_formulae}). 
Because the general form (namely 3-integral) of Jeans equations are not closed, 
instead of solving it with the above-mentioned simplifications, 
we use the two independent Jeans equations, $R$- and $z$-directional, 
as two discriminators of the consistency between gravitational potential models and kinematic data.
In order to (1) reduce the impact of various kinematic substructures in the Galactic disk, 
as well as (2) to avoid the numerical artifacts caused by numerical differentiation in Jeans-equations calculation
given the limited spatial resolutions of the observational data,
we apply the analytic form for $\sigma_R(R,z)$ and $\sigma_z(R,z)$ proposed by \citet{binney_galactic_2014},
and successfully extend it to the azimuthal components $\sigma_\theta(R,z)$ and $V_\theta(R,z)$.
Our comprehensive tests consistently point to the conclusion:  
Whereas the current kinematic data, 
with the precision and accuracy powered by Gaia DR2,
is able to reject the MOG model (let alone the Newtonian baryon-only model; 
\QesComm{adopting the baryonic mass distribution priorly best-fitted in the DM paradigm}),
the MOND model is still not rejected, and behaves as good as the DM model.
This is surprising, because  
\QesComm{whereas the fiducial DM model we adopt was carefully pre-fitted 
with all available Galactic kinematic data 
and in fact has been kept improving elaborately by researchers during past decades 
(see \S\ref{sec:fiducialMWmodel} and the references therein)}, 
there is no free parameter at all in the MOND model (no bother of fitting), 
and the parameters of the baryonic mass model 
are actually fine-tuned in the DM context.

This paper is organized as follows. 
In Section~\ref{sec:jeans}, we describe the complete form of Jeans equations for axisymmetric systems,
and propose the two measures $T_R$ and $T_z$.
In Section~\ref{sec:model}, we give the fiducial mass distribution model of the MW used in this work,
with best-fit model parameters in the DM context (\S\ref{sec:fiducialMWmodel}),
and describe two alternative gravitational potential models, 
namely quasi-linear MOND (\S\ref{sec:qumond}) and MOG (\S\ref{sec:mog}).
In Section~\ref{sec:data}, we introduce the data we employ; in particular, in \S\ref{sec:our_kinematics_formulae}
we describe our further analysis of the 3D velocity data of \citet{huang_mapping_2020-1},
and present our best-fit formulae for the spatial distributions, 
namely $\sigma_R(R,z)$, $\sigma_z(R,z)$, $\sigma_\theta(R,z)$ and $V_\theta(R,z)$.  
In Section~\ref{sec:results}, we present the results of comprehensive tests on the gravitational potential models,
particularly the $T_R$ and $T_z$ tests using different tracer populations 
with various density profiles of tracers assumed (\S\ref{sec:results_JETs} and \S\ref{sec:results_common_tracers});
in \S\ref{sec:effective_equivalence} we discuss the physical implication as well as its practical application
of our main result.
In addition, in the Appendix 
we present the results using 
a different parameterization of the Galactic mass model and corresponding kinematic data,
which are consistent with the results in the main text.
Section~\ref{sec:summary} summarizes the paper.
\QesComm{
Throughout the paper, we adopt a Galactocentric cylindrical system, 
with $R$ being the projected Galactocentric distance, increasing radially outwards, 
$\theta$ toward the Galactic rotation direction, and $z$ in the direction of north Galactic pole.
}.

\section{Two discriminators in terms of the three-integral Jeans equations}\label{sec:jeans}

Rotation curves, which involve rotational velocities (i.e., in the azimuthal direction) only,
and are conventionally measured in the galactic mid-plane only,
are one-dimensional:
reflecting the azimuthally averaged $R$-directional acceleration; i.e., 
$V^2_\mathrm{c}(R) /R = \partial \Phi / \partial R(R;z=0)$.

Most previous applications of Jeans equations, as described in the Introduction,
assumed two-integral dynamics 
and even additional simplifications,
which are not consistent with the observed kinematic data of the WM, the subject of the present study.

Our own Galaxy provides 3-dimensional data, i.e., the 3-directional components of velocity-related quantities (see \S\ref{sec:data}).
Moreover, it enable us to test gravitational potential models \textit{at different spatial locations ($R, z$)}, 
or even at 3-dimensional locations ($R, \theta, z$) in the future.
This is in stark contrast with external galaxies,
where only vertically-averaged quantities are available, 
such as observed radial $\sigma_z$ profiles (e.g., the DiskMass project;  
see \citealt{angus_mass_2015}).
\footnote{\,The $\sigma_z$ profiles mean $\sigma_z(R)$ where $\sigma_z$ is averaged or integrated over the $z$ direction, 
	similar to the form of RCs $V_\mathrm{c}(R)$. Likewise, in radial profiles of line-of-sight (LOS) \sigmastar\  
	(namely $\sigma_{\rm los}(R)$; e.g. \citealt{kipper_stellar_2016}), $\sigma_{\rm los}$ is averaged or integrated along the line of sight.}
The point is, the set of kinematic data (\sigmastar\ and $\bar{v}_\theta$\/) at {\em every} ($R, z$) location 
can be regarded as an {\it independent} constraint to the gravitational models through Jeans equations, 
and thus the more data points---\textit{particularly those at relatively large $z$}---the better the models get constrained.

To test gravitational models comprehensively, with 3-dimensional kinematic quantities 
(namely their $R$-, $z$-, and $\theta$-directional components)
and at different $(R,z)$ locations, 
we invoke the complete form of Jeans equations.
For a steady-state collisionless gravitational system, 
Jeans equations relate the gravitational field of the system to 
the density and kinematic qualities of a certain tracer population (\S4.8 of \citealt{binney_galactic_2008}). 
We write the equations using the notations of \citet{kipper_stellar_2016}. 
Because the mass models we use are axisymmetric (see \S\ref{sec:fiducialMWmodel}), 
the two cross-term components
of the velocity dispersion tensor are zero, $\sigma_{R\theta} = \sigma_{\theta z} = 0$. 
Thus, the complete Jeans equations can be written as two independent equations in cylindrical coordinates:
\begin{equation}
	\label{eq:Jeans_R}
	\frac{\partial(\rho \sigma_R^2)}{\partial R} + \left(\frac{1-k_\theta}{R}
	+\frac{\partial \kappa}{\partial z}\right)\rho \sigma_R^2 + \kappa
	\frac{\partial(\rho \sigma_R^2)}{\partial z} - \rho \frac{V_\theta^2}{R}= -\rho
	\frac{\partial \Phi}{\partial R},
\end{equation}
\begin{equation}
	\label{eq:Jeans_z}
	\frac{\partial(\rho \sigma_z^2)}{\partial z} + \left(\frac{\xi}{R}+
	\frac{\partial \xi}{\partial R}\right)\rho \sigma_z^2 + \xi
	\frac{\partial(\rho \sigma_z^2)}{\partial R} = -\rho
	\frac{\partial \Phi}{\partial z} ~ ,
\end{equation}
where
$\kappa = \frac{1}{2}\tan(2\alpha)(1-k_z)$,
$\xi = \kappa/k_z$,
and $V_\theta \equiv \bar{v}_\theta$,  the averaged azimuthal velocity of tracers at every location.
The parameter $\alpha$ is the tilt angle of the velocity ellipsoid, i.e.,
the angle by which the ellipsoid's longest axis at every position is tilted with respect to the galactic-disk plane.
The other two parameters, $k_z$ and $k_\theta$, are the axial ratios of the ellipsoid:
$k_z = \sigma_z^2/\sigma_R^2$
and
$k_\theta = \sigma_\theta^2/\sigma_R^2$ . 
Note that $\rho$ in Jeans equations is tracer's density,
while $\Phi$ is the total gravitational potential contributed by all components of the system.

Given observed kinematic data, a right gravitational model or theory 
should satisfy the two Jeans equations everywhere throughout the MW. 
As mentioned, because the equations are not closed, 
we define two measures as follows,
\begin{equation}
	\label{eq:TR}
    T_R = -\frac{1}{\rho} \left\{
        \frac{\partial(\rho \sigma_R^2)}{\partial R} + \left(\frac{1-k_\theta}{R}
	    +\frac{\partial \kappa}{\partial z}\right)\rho \sigma_R^2 + \kappa
	    \frac{\partial(\rho \sigma_R^2)}{\partial z} - \rho \frac{V_\theta^2}{R}\right\} ~ ,
\end{equation}
and
\begin{equation}
	\label{eq:Tz}
    T_z = -\frac{1}{\rho} \left\{
        \frac{\partial(\rho \sigma_z^2)}{\partial z} + \left(\frac{\xi}{R}+
	    \frac{\partial \xi}{\partial R}\right)\rho \sigma_z^2 + \xi
	    \frac{\partial(\rho \sigma_z^2)}{\partial R}
    \right\} ~ .
\end{equation}
According to the Jeans equations (Equations~\ref{eq:Jeans_R} and \ref{eq:Jeans_z}),
a correct gravitational model ($\Phi$) should satisfy
\begin{equation}
\label{eq:TR_test}
T_R = \frac{\partial \Phi}{\partial R}
\end{equation} 
and 
\begin{equation}
\label{eq:Tz_test}
T_z = 	\frac{\partial \Phi}{\partial z}
\end{equation} 
everywhere throughout the MW. 
We call the above two criteria ``$T_R$ test'' and ``$T_z$ test'', respectively.
We will see (\S\ref{sec:results_common_tracers}), 
the discriminating power of $T_R$ test comes from the fact that it is fairly insensitive to 
the choice of tracer's density profile (namely the common prescriptions for galactic components),
while the merit of $T_z$ test is instead its sensitivity to  tracer's density profile.

The measure $T_R$, in fact, is the observed $R$-directional acceleration,
calculated from $\sigma_R$, $\sigma_z$ (through $\kappa$), 
$\sigma_\theta$ (through $k_\theta$), $V_\theta$ and tilt angle $\alpha$ (through $\kappa$), 
as well as tracer's density profile $\rho(R,z)$.
Thus $T_R$ test means that the observed $R$-directional acceleration equals to 
the radial gradient of gravitational potential at any locations.
It can be regarded as a generalized rotation-curve test, on and off the galactic mid-plane (\citealt{chrobakova_gaia-dr2_2020}).

Likewise, the measure $T_z$ is the observed $z$-directional acceleration,
calculated from $\sigma_z$ and tilt angle $\alpha$ (through $\xi$), 
as well as tracer's density profile $\rho(R,z)$.
$T_z$ test means that the observed $z$-directional acceleration equals to 
the vertical potential gradient at any locations.
In testing the vertical characteristics of gravitational models,
\QesComm{
	$T_z$ is more universal and accurate (i.e., without additional simplifications) than 
	the old $K_z$ (``vertical force'') method as mentioned in the Introduction 
	(see also \S3.3 of \citealt{Read_2014_JPhG_41}, and \S\ref{sec:results_JETs} below), 
	that is generally either partially one-dimensional (e.g., neglecting the tilt term in vertical Jeans equation; 
	e.g., \citealt{McGaugh_2016_RCgradient}),
	or even completely one-dimensional
	(in both vertical Jeans and Poisson equations; e.g., \citealt{Lisanti_2019_PRD}).
}	
	
\QesComm{
Note that we deliberately use a different terminology ``$T_z$'' (as well as ``$T_R$'') rather than the old ``$K_z$'',
	in order to avoid any possible prejudice resulting from the simplified use of the ``$K_z$'' method prevailing in the literature,
	and to stress that our two measures by definition are \textit{observed} vertical and radial \textit{kinematic accelerations}
	calculated from tracers' density profile and kinematic data.
	By definition $K_z$ is vertical \textit{field strength}, namely the negative of gradient, 
	of (theoretical models of) gravitational potential.
	Because of the same consideration, in this paper we often use the words ``acceleration'' vs. ``field strength, force or gradient'' differently.   
}

\section{Mass distribution (Potential) models of the Milky Way}\label{sec:model}
In this work we focus on the global potential field of the MW, particularly the outer part where the circular velocity and velocity dispersion are dominated by the supposed DM, we therefore choose to ignore kinematic substructures of stars, 
and non-axisymmetric structures (e.g., bars and spiral arms) that are dynamically important mainly in the inner part. 
Specifically, we use and compare axisymmetric mass models of the MW throughout the paper.

For all the models, we implement a light-weight {\tt C} program to solve the axisymmetric potential 
\QesComm{on a $1280^3$ grid  
using the direct sum method \citep[][]{binney_galactic_2008}.
The grid is equally divided into cells, and every cell physically corresponds to 
a spatial size \mbox{60\,pc} on a side. 
}
We have checked the numerical convergence and verified our results, 
using {\tt FreeFem++} \citep{hecht_new_2012}, 
a popular software solving partial differential equations with the finite-element method (FEM),
which achieves both high spatial resolution and high precision.

\subsection{Fiducial model of the Galactic mass distribution}
\label{sec:fiducialMWmodel}
The fiducial mass model we use in the main text
is the one prescribed by \citet{wangMilkyWayTotal2022}.
It adopts the mass distribution profile formulae and basic structural parameter values from
the best-fit main model of  \citet{mcmillan_mass_2017} for the bulge, stellar disks and interstellar medium disks,
and the Zhao's (\citealt{zhaoAnalyticalModelsGalactic1996b}) profile for the DM halo.
The density values (namely the normalization of the aforementioned profiles),
as well as the scale lengths of thin and thick stellar disks
and the other parameters of the DM halo, are constrained by \citet{wangMilkyWayTotal2022} with latest observations
powered by 
Gaia DR2 \citep{collaborationGaiaDataRelease2018} and Gaia EDR3 \citep{collaborationGaiaEarlyData2021}. 
We briefly summarize the details of every components below.

The bulge's density profile is
\begin{equation}\label{eq:bulge}
  \rho_\mathrm{b}=\frac{\rho_{0,\mathrm{b}}}{(1+\frac{r^\prime}{r_0})^\alpha}\;
  \textrm{exp}\left[-\left(\frac{r^\prime}{r_{\mathrm{cut}}}\right)^2\right],
\end{equation}
and, in cylindrical coordinates,
\begin{equation}
  r^\prime = \sqrt{R^2 + \left(\frac{z}{q}\right)^2},
\end{equation}
with
$\rho_{0,\mathrm{b}} = 9.5\times10^{10} {M_\odot \rm kpc^{-3}}$, $\alpha=1.8$, $r_0=0.075$ kpc, $r_\mathrm{cut}=2.1$ kpc,
and axis ratio $q=0.5$.

The stellar disks of the Milky Way are usually considered to be divided into two components:
the thin disk and thick disk.
Their mass distributions follow the following form
\begin{equation}\label{eq:disk}
  \rho_\mathrm{d}(R,z)=\frac{\Sigma_{0}}{2z_\mathrm{d}}\;\textrm{exp}\left(-\frac{\mid
      z\mid}{z_\mathrm{d}}-\frac{R}{R_\mathrm{d}}\right),
\end{equation}
with corresponding scale height $z_\mathrm{d}$, scale length $R_\mathrm{d}$, and central surface
density $\Sigma_{0}$.

The interstellar medium of the Milky Way includes two components:
the \HI\ and molecular $\mathrm H_2$ gas disks.
These disks follow the density law
\begin{equation}\label{eq:gasdisc-M17}
  \rho_\mathrm{g}(R,z)=\frac{\Sigma_{0}}{4z_\mathrm{d}}\;
  \exp \left(-\frac{R_{\rm m}}{R}-\frac{R}{R_\mathrm{d}}\right)\;
  {\mathrm{sech}}^2(z/2z_\mathrm{d}),
\end{equation}
with $R_{\rm m}$ being 
\QesComm{
	the associated scalelength of the central hole.
	The actual width of the hole is determined by both $R_\mathrm{d}$ and $R_\mathrm{m}$,
	with the maximum surface density (i.e., the rim of the hole) being 
	at $R = \sqrt{R_\mathrm{d} R_\mathrm{m}}$.
}
The parameters of the stellar and gas disks are listed in Table \ref{tab:model-W21}.

The DM halo is described by the Zhao's profile,
\begin{equation}
\rho_\mathrm{h}(r)=\rho_{0, \mathrm{h}}\left(\frac{r}{r_{\mathrm{h}}}\right)^{-\gamma}\left[1+\left(\frac{r}{r_{\mathrm{h}}}\right)^{\alpha}\right]^{(\gamma-\beta) / \alpha},
\end{equation}
where the full set of three free parameters $(\alpha, \beta, \gamma)$ can be calculated analytically. In this paper, we adopt the best derived value $(\alpha, \beta, \gamma)=(1.19, 2.95, 0.95)$ (see Table 2 of \citealt{wangMilkyWayTotal2022}). The Zhao's profile is more flexible than the widely used NFW \citep{navarro_structure_1996} profile, and will reduce to the normal NFW formula in the case of $(\alpha, \beta, \gamma)=(1,3,1)$.
The remaining halo parameters are as follows: 
$\rho_{0,\mathrm{h}} = 1.55\times 10^7 {M_\odot \rm kpc^{-3}}$, $r_{\rm h}=11.75$ kpc, and $q=0.95$.

In this study (except in the Appendix), we use the distance from the Sun to the Galactic center 
 $R_\odot=8.122~\rm kpc$ \citep{abuterDetectionGravitationalRedshift2018},
and a nominal circular velocity  $v_\odot \simeq 229.0~\rm km\,s^{-1}$ at the radius of the Sun \citep[e.g.][]{eilers_circular_2019}.
The fiducial model of the Galactic mass distribution was built under the same $R_\odot$ and $v_\odot$ constants,
i.e. the same as \citet{eilers_circular_2019} (J.Wang 2022, private communication).

We have explored other parameterizations of the Galactic mass distribution (as well as other kinematic data),
including those under other sets of the solar position and velocity values ($R_\odot$ and  $v_\odot$),
and found that our conclusions remain intact.
Such an examination, performed under the legacy $R_\odot$ and  $v_\odot$ values, is presented in the Appendix.
\\

\QesComm{
Finally, because the model parameters of the above Galactic components were constrained in the DM context,
for fair comparison between DM and modified-gravity models we need to make clear 
to what degree the data used in constraining the fiducial mass model (by \citealt{mcmillan_mass_2017} and \citealt{wangMilkyWayTotal2022})
overlap the data we use here to discriminate gravitational models.
Here we summarize the data that were already used to fit the fiducial mass model,
and list the overlapped parts with the data used in the present study.
\citet{mcmillan_mass_2017} used various rotation-curve data, solar velocity (to constrain $v_\odot$),
vertical-force data at $|z| = 1.1$ kpc and $R = R_\odot$ of \citet{KG91_Kz_data},
 and the upper limit of the total mass within the MW's inner 50 kpc according to \citet{WE99_M50_data}.
\citet{WE99_M50_data} based their estimate on the distance and velocity data of 27 objects in the outer Galaxy 
(satellite galaxies and globular clusters at $R > 20$ kpc).
\citet{wangMilkyWayTotal2022} used the rotation-curve data of \citet{eilers_circular_2019},
the vertical-force data at $|z| = 1.1$ kpc and  $4 \lesssim R  \lesssim 9$ kpc,
$K_{z,\mathrm{1.1kpc}}(R)$, derived by \citet{Bovy_2013_Kz_data} based on G-type dwarf stars from SDSS/SEGUE survey
(see also \S\ref{sec:results_vexing_at_small-z}), 
and kinematic data of globular clusters.
In relation to the data used in the present study (see \S\ref{sec:data}),
(1) the rotation-curve data, concerning radial accelerations in the Galactic plane, are essentially overlapped
(particularly the best data obtained by \citealt{eilers_circular_2019});
(2) the radial accelerations off the Galactic plane (namely rotation curves at $|z|>0$) probed by our data
are not available in either \citet{mcmillan_mass_2017} or \citet{wangMilkyWayTotal2022};
(3) as to the data concerning vertical accelerations (e.g., the so-called ``vertical force'' $K_z$ data),
in effect there is overlap to a certain degree, but the vertical accelerations probed by our data are not limited at $|z| = 1.1$ kpc;
(4) the data of satellite galaxies and globular clusters used by 
\citet{mcmillan_mass_2017} and \citet{wangMilkyWayTotal2022} are completely irrelevant to our data.
}

\begin{table}
  \centering
  \caption{Disk parameters for the Milky Way mass model we use. }
  \label{tab:model-W21}
  \begin{tabular}{lcccc}
  \hline \hline
  ~                         & Thin    & Thick    & \HI    & $\mathrm{H}_2$ \\ \hline
  $\Sigma_0 [\!\msun \pc^{-2}]$ & 1003.12  & 167.93   & 53.1 & 2179.5         \\
  $R_{\mathrm d} [\!\kpc]$       & 2.42     & 3.17     & 7.0    & 1.5            \\
  $z_{\rm d} [\!\kpc]$           & 0.3     & 0.9      & 0.085  & 0.045          \\
  $R_{\rm m} [\!\kpc]$           & -       & -        & 4.0    & 12.0           \\ \hline
  \end{tabular}
\end{table}

\subsection{Quasi-linear MOND}
\label{sec:qumond}
QUMOND is the quasi-linear realization \citep[][]{milgrom_quasi-linear_2010} 
of the MOND theory \citep[][]{milgrom_modification_1983}.
MOND was initially proposed to explain the flat rotation curves of galaxies without DM. 
We refer the reader to recent reviews (such as \citealt{famaey_modified_2012} and \citealt{Banik_Zhao_2022_review})  
for detailed and lucid descriptions. 
In this work, essentially we treat QUMOND as a gravitational potential model 
rather than a ``modified gravity or dynamics'' \textit{theory};
i.e., we employ it in the fashion of $\rho_{\rm b} + \rho_{\rm pdm}$, 
with $\rho_{\rm pdm}$ as an alternative of popular DM halos.
\QesComm{Here, ``pdm'' (or in capital letters) means ``phantom dark matter'',
	a term coined to reflect that  
	this MOND effect---such a virtual (phantom) stuff---would be 
	interpreted by a Newtonist as a DM halo (see below).}  
We calculate the QUMOND potential with the baryonic mass density profile prescribed in the fiducial mass model.

The MOND acceleration was originally written in the following way (the \citealt{milgrom_modification_1983} formula):
\begin{equation}
\label{eq:milgrom_1983_formula}
\mu\left(\frac{g}{a_{0}}\right) \mathbf{g}=\gN
\end{equation}
where $\mu(x)$ is an interpolating function, and
\begin{equation}
\mu(x) \rightarrow 1 \text { for } x \gg 1 \text { and } \mu(x) \rightarrow x \text { for } x \ll 1  ~ .
\end{equation}
Here $\gN$ is the Newtonian gravitational acceleration $\gN = -\grad{\Phi_{\rm N}}$,
and $\PhiN$ is the Newtonian potential:
\begin{equation}
\nabla^2 \PhiN = 4 \pi G \, \rhob,
\label{eq:step1}
\end{equation}
with \rhob\ being the baryonic matter density. 
In terms of the simple \citealt{milgrom_modification_1983} formula (Equation~\ref{eq:milgrom_1983_formula}), however, 
the acceleration field $\mathbf{g}$ is not derivable from a scalar potential, 
and consequently there is no conserved momentum.

QUMOND, just like its cousin AQUAL 
(aquadratic Lagrangian formulation of MOND, \citealt{bekensteinDoesMissingMass1984}), 
is a complete theory that is self-consistently derivable from an modified Newtonian gravitational action 
(see \citealt{famaey_modified_2012} for the detail). 
QUMOND has the following Poisson equation:
\begin{equation}
\nabla^2 \Phi =
\nabla \cdot \left[ \nu(\frac{\gNabs}{a_0}) \, {\mathbf\grad}\PhiN \right] ~ ,
\label{eq:qumond}
\end{equation}
where scalar $\Phi$ is the QUMOND gravitational potential, and $\nu(y)$ is an interpolating function.
The function $\nu(y)$ is related to the above $\mu(x)$ by $  \nu(y)  = 1 /\mu(x) $
and $ y = x \mu(x) $.
We can define $\nut(y) = \nu(y) - 1$, then Equation~(\ref{eq:qumond}) leads to
\begin{equation}\label{eq:qumond2}
\nabla^2 \Phi  =
\nabla \cdot \left[ {\mathbf\grad}\PhiN + \nut(\frac{\gNabs}{a_0}) \, {\mathbf\grad}\PhiN \right],
\end{equation}
or
\begin{equation}
        \nabla^2 \Phi =  4 \pi G ( \rhob + \rhoph ) ~ .
\label{eq:step3}
\end{equation}  

Equation~\ref{eq:step3} reveals the merit of QUMOND: 
the gravitational potential can be ascribed formally to two matter sources in terms of normal Poisson equation,
the baryonic matter and the aforementioned PDM.  
\QesComm{From a mathematical point of view, the PDM density \rhoph\ is conceptually equivalent 
	to the density of ``DM halos'' (but with totally different physical content);
see \S\ref{sec:results_extra_mass_gravity},
also \citealt{milgrom_quasi-linear_2010} and Section~6.1.3 of \citealt{famaey_modified_2012}.}
Accordingly, there is a striking technical advantage 
(e.g., compared with AQUAL that involves a non-linear generalization of Poisson), 
which is obvious:
QUMOND involves solving only \textit{linear} differential equations (namely the normal Poisson equation).
Thus, \textit{all} the well-developed algorithms (e.g., Tree-PM) and codes 
(e.g., {\tt Gadget} of \citealt{springel_cosmological_2005}) for Newtonian N-body
numerical calculations and simulations
are still usable in QUMOND.
\footnote{In the literature, there was a claim that due to the non-linearity of MOND,
	the Poisson solvers that are not based on grids/meshes, such as tree-codes,
	cannot be used (e.g., \S2.1 of \citealt{angus_cosmological_2013}). This is not necessarily true for QUMOND, because
	one can build a temporary grid to implement Equation~(\ref{eq:step2}), calculating PDM density from Newtonian potential,
	which is not difficult technically (\S6.1.3 of \citealt{famaey_modified_2012}). 
}

In practice, given baryonic \rhob\ or \PhiN, 
\rhoph\ is calculated straightforward as follows,
\begin{equation}
\rhoph = \frac{1}{4\pi G} \, \nabla \cdot \left(  \nut(\frac{\gNabs}{a_0})  \, {\mathbf\grad}\PhiN \right)    ~ .
\label{eq:step2}
\end{equation}
Correspondingly, we can trivially define a scalar $\phiph$ as the PDM potential,
\begin{equation}
\nabla^2 \phiph  =
 \nabla \cdot \left(  \nut(\frac{\gNabs}{a_0}) \, {\mathbf\grad}\PhiN \right)   = 4 \pi G \, \rhoph ~ ,
\end{equation}
then the QUMOND potential can be written as 
$\Phi = \PhiN + \phiph$\,.

In this work, the critical acceleration constant is held fixed to be the commonly used value 
 $a_0 = 1.2\times 10^{-10} \,\mathrm{m\,s}^{-2}$ \citep{Banik_Zhao_2022_review}. 
The simple formula of $\nu(y)$ is adopted \citep{famaey_modified_2012}:
\begin{equation}
    \nu(y) = \frac{1}{2} \sqrt{1 + \frac{4}{y}} + \frac{1}{2}  ~.
\end{equation}
That is, there is no free parameter at all in the QUMOND formula that we use in this study.

Note that in this work we have not taken into account the so-called ``external field effect'' (EFE) of MOND.
EFE is a general characteristic of MOND (particularly its modified-gravity theories such as QUMOND),
because MOND depends on the {\it total} acceleration with respect to some pre-defined (inertial) frames.
But EFE does not necessarily exist in specific MOND theories (see \S4.6 of \citealt{milgromMONDLawsGalactic2014}),
particularly in modified-inertia theories of MOND (see \citealt{milgromMONDParticularlyModified2011}).
Thus, in this work, we only practically use QUMOND as a practical (effective) formula to
calculate the MOND potential of the MW baryons, and refrain from accounting for the subtlety of EFE.
Anyway, practically, the gravitational strength of the external field around the MW is 
reasonably estimated to be 0.01--0.03\,$a_0$ \citep{wu_milky_2008},
which is $\sim 10^2$ times smaller than 
the Newtonian gravitational strength at the $(R,z)$ locations considered in this work; i.e., 
the EFE is negligible for our purpose.
In addition, mention in passing that, by defining $\mathbf{g} = - \nabla \Phi$, 
the complete MOND theories so far (such as QUMOND and AQUAL) assume the gravitational vector field 
is still curl-less; 
in contrast, the gravitational or acceleration field $\mathbf{g}$ defined in the pristine \citet{milgrom_modification_1983} formula
(Equation~\ref{eq:milgrom_1983_formula})  is curled.

\subsection{Moffat's MOG}
\label{sec:mog}
We also test another alternative to DM, Modified Gravity 
(MOG; e.g., \citealt{moffatScalarTensorVector2006,moffatMOGWeakField2013}), 
which is a covariant modification of Einstein gravity.  
Simply put, MOG adds two additional scalar fields and one vector field 
to explain the dynamics of astronomical systems based on the distribution of baryonic matter. 

In the weak field approximation (e.g., in the MW), 
the effective potential for an extended distribution of baryonic matter ($\rho$) in MOG
is as follows:
\begin{equation}
    \Phi(\boldsymbol{x}) = -G_\infty \left[ \int{ \frac{\rho(\boldsymbol{x}')}{|\boldsymbol{x}-\boldsymbol{x}'|}
    \left(
    1 - \frac{G_\infty-G_{\rm N}}{G_\infty}{\rm e}^{-\mu|\boldsymbol{x}-\boldsymbol{x}'|}
    \right) {\rm d}^3 \boldsymbol{x}'
    } \right].
\end{equation}
with $G_\infty=(1+\alpha)G_{\rm N}$ being the modified gravitational constant. 
In this work, the two universal constants are held fixed to be 
$\alpha=8.89$ and $\mu=0.042\,\rm kpc^{-1}$,
which are best fitted with the rotation-curve data of external galaxies by \citet{moffatMOGWeakField2013}.

\section{Data} \label{sec:data}
We use recent kinematic observations, including the rotation curve and 3-dimensional velocity dispersion, of the MW to test 
the gravitational models. 
We only include data at $R>4\kpc$ to avoid the complexity in the central region of the MW.
Besides the data collected from the literature,
we analyze and fit the spatial distributions along $R$ and $z$ directions 
of $\sigma_R$, $\sigma_z$, $\sigma_\theta$ and $V_\theta$ (namely the mean azimuthal velocity, see \S\ref{sec:jeans}).
We basically follow the methodology of \citet{binney_galactic_2014}, 
except for an additional innovation that we also give well-parameterized formulae 
for $\sigma_\theta(R,z)$ and $V_\theta(R,z)$,
which are described below (\S\ref{sec:our_kinematics_formulae}).

Our own Galaxy is remarkable in testing gravitational models. 
There are already plenty of kinematic observations of both RC and \sigmastar\ (as well as $\bar{v}_\theta$). 
Moreover, although on the one hand our position inside the Galactic disk weakens the ability to measure the RC in the outer Galaxy, 
on the other hand it allows a three-dimensional measurement of the position and velocity of individual stars, 
particularly of those in the $z$ direction far into the halo.

\subsection{Spatial-distribution formulae for 3-dimensional kinematics based on Gaia DR2} \label{sec:our_kinematics_formulae}

We analyze the three-dimensional velocity data of
the {\tt LAMOST and Gaia red clump sample}
complied by \citet{huang_mapping_2020-1}.
This sample, consisting of $\approx137,000$ red clump stars (as the tracer population in this work),
has a good coverage of the Galactic disk of $4\leq R \leq 16$ kpc and $|z|\leq4$ kpc.

In order to reduce the impact of particular structures in the Galactic disk 
(e.g., stellar streams of various origins; \citealt{gaia_aa_2018_616_a11}),
we fit the velocity dispersion, $\sigma_R$, $\sigma_\theta$, and $\sigma_z$,
to the smooth analytic forms with respect to $R$ and $z$
given by \citet[][particularly cf. their Tables 2 and 3]{binney_galactic_2014},
and thus acquire the ``macro'' (namely spatially coarse-grained) kinematics.
For the same reason, we make no efforts in distinguishing different stellar groups, 
although we appreciate the difference in the kinematics of stars with different age and metallicity \citep[][]{huang_mapping_2020-1}.

In fact, 
we have tried using only the red clump stars in the thin disk 
(numbering $\approx$116,000; according to the [Fe/H]--[$\alpha$/Fe] criterion by \citealt{huang_mapping_2020-1}) 
as the tracer population, which would enable us to have a better constraint on the density profile of the tracers
(cf. \S\ref{sec:results_common_tracers}),
e.g., by simply adopting the geometrical thin-disk component in the fiducial mass model
as the tracer's density profile.
But, it turns out that if we do so, many spatial bins at $|z| > 1$~kpc
have not sufficient stars to fit the $v_\theta$ probability distribution (see below),
and thus disable us to perform the $T_R$ tests for those spatial locations.
Because $T_R$ is the important and robust measure to test the gravitational models 
(see \S\ref{sec:results_JETs} and \S\ref{sec:results_common_tracers}),
we base our main results of this work on the entire red clump sample of \citet{huang_mapping_2020-1},
and for safety we test our results by using three schemes of density profile for the tracers.
Besides, the results based on the thin-disk-only red clump stars 
are consistent with those based on the entire sample (see \S\ref{sec:results_common_tracers}).

\citet{binney_galactic_2014} presented parameterized formulae for the spatial distributions of 
the two meridional-plane components of stellar velocity dispersions (i.e., velocity ellipsoid),
namely $\sigma_1(R,z)$ and $\sigma_3(R,z)$, as follows (their Equation~4):
\begin{equation}
\label{eq:sigma-binney}
\sigma(R, z)=\sigma_{0} a_{1} \exp \left[-a_{2}\left(R / R_\odot-1\right)\right]\left[1+\left(a_{3} z / R\right)^{2}\right]^{a_{4}}  ~.
\end{equation}
The above functional form comes out of physical intuition as well as their trial and error,
and work well in practice. 
Here we adopt the same formulae for $\sigma_R$ and $\sigma_z$, 
and set the parameters (see the constants in Table~\ref{tab:ourParameters-KinematicFormulae}) to be free 
and constrained by our data.
Formally there seems a difference 
in that the formulae of \citet{binney_galactic_2014} are for the two principal velocity dispersions 
and here for those along the $R$ and $z$ directions.
But in essence this is not a problem (considering the semi-empirical nature of the formulae), 
and has been verified by our experiment.
Rather, this is partly the reason that we allow our best-fit constants 
can be different to some degree from those of \citet{binney_galactic_2014}.

\citet{binney_galactic_2014} presented a novel fitting recipe (see their Equations \mbox{7 \& 8}) 
to model the distributions of the azimuthal velocities ($v_\theta$) of the tracers on every spatial location, 
i.e., for their \emph{every ($R,z$) bin}; 
it is well-known, as the \textit{asymmetric drift} phenomenon, 
that the $v_\theta$ distributions are highly non-Gaussian.  
The \citet{binney_galactic_2014} distribution function takes a form of \textit{sigma-varying} Gaussian,
i.e., with different $\sigma$ (dispersion) for different $v_\theta$, for the sample of tracer stars in a spatial bin;
the fitting is extremely good when applied to observed data.

We make a further innovation on the shoulder of \citet{binney_galactic_2014}
out of our exploration:
The spatial distributions of the mean azimuthal velocity ($V_\theta$) 
and its corresponding $\sigma_\theta$ 
(derived by the \citealt{binney_galactic_2014} methodology, as described in the above paragraph),
i.e., distributions over a range of spatial bins,  
can be well fitted by the formula of Equation~\ref{eq:sigma-binney} also.
The possibility of such an innovation was actually discussed by \citet[][see their \S4.1]{binney_galactic_2014}, 
although the RAVE data they used only cover a small region within $\sim$2~kpc of the Sun.  
We now have a sample of $V_\theta$ and $\sigma_\theta$ data  
with larger coverage in the $R-z$ plane than \citet{binney_galactic_2014}, 
which exhibit apparent trends of $V_\theta$ and $\sigma_\theta$ over large spatial scales enabling us to conduct such an exploration.

Following the methodology described in the above three paragraphs,
we calculate the quantities $\sigma_R, \sigma_z, \sigma_\theta$ and $V_\theta$ for every spatial bins,
and then fit their spatial distribution with Equation~\ref{eq:sigma-binney}.
The $\sigma_{0}$ in the equation is fixed to be $30~\kms$. 
Our best-fit parameters are listed in Table~\ref{tab:ourParameters-KinematicFormulae}.

Besides the merit of the well-parameterized analytic formulae of $\sigma_R(R,z)$, $\sigma_z(R,z)$,
$\sigma_\theta(R,z)$ and $V_\theta(R,z)$ \textit{per se},
the analyticity of $\sigma_R(R,z)$ and $\sigma_z(R,z)$ leads to a great advantage in calculating $T_R$ and $T_z$:
derive the partial derivatives analytically (such as $\partial \sigma_R / \partial R$, 
$\partial \kappa / \partial z$, $\partial \sigma_z / \partial z$, etc.), 
free of the technical difficulties in calculating those partial derivatives numerically instead 
(e.g., ``huge'' errors in such numerical implementation given the spatial resolutions so far; 
see, e.g., \citealt{chrobakova_gaia-dr2_2020}).
Again, we would like to stress that our major motivation of using these spatial-distribution formulae
is to reduce the astrophysical ``impurities'' such as kinematic substructures.

Regrading the spatial binning of the data, 
generally we divide the entire space ($4<R<16$ kpc and $-4<z<4$ kpc) 
into bins of $\Delta R = 0.2$ kpc and $\Delta z = 0.05$ kpc. 
We use the bins with more than 10 objects 
to fit Equation~\ref{eq:sigma-binney} for $\sigma_R(R,z)$, $\sigma_z(R,z)$, and $V_\theta(R,z)$. 
As for $\sigma_\theta$, in order to get relatively reliable fitting 
to its statistical distribution within a specific bin 
(Equation~7 of \citealt{binney_galactic_2014}), 
we only employ the bins with more than 50 objects,
derive their $\sigma_\theta$, and fit Equation~\ref{eq:sigma-binney} 
for $\sigma_\theta(R,z)$. 
By and large, the spatial bins over the ``continuous'' space of $6< R < 12$ kpc and $-2.5 < z < 2.5$ kpc
have  $\sigma_\theta$ measurements.
\QesComm{
Thus, the reliable $R$ range 
for applying the best-fit spatial distributions (Equation~\ref{eq:sigma-binney})
is $6< R <12$ kpc (without poor fitting on the boundaries because of 
abundant data at $R<6$ and $R>12$ kpc);
the reliable $|z|$ (the distance to the Galactic mid-plane) range
is conservatively deemed to be 
from the resolution limit (see \S\ref{sec:results_vexing_at_small-z}) to $|z| = 2$ kpc.
}

Regarding the measurement uncertainties of the velocity dispersion values in the spatial bins,
the 1-$\sigma$ statistical errors in $\sigma_R$ are $<2.8$ \kms,
those in $\sigma_z$ are $<2.5$ \kms,
and those in  $\sigma_\theta$, $<3.4$ \kms;
the mean error in any one of the three quantities is 0.5~\kms.
The above quoted errors already include the effect of systematic errors in distance estimation 
on the derived kinematic quantities, 
because the uncertainties of the 3D velocities given by \citet{huang_mapping_2020-1}
have accounted for all kinds of error sources by Monte Carlo simulation.
In fact, the total measurement uncertainty in distance is 5--10\% 
(see \S5.2 of \citealt{huang_mapping_2020-1}),
to which the contribution of systematic bias is minor by virtue of the power of Gaia.
This is totally different from the situation prior to the Gaia era (cf. \S5.3 of \citealt{binney_galactic_2014}).

The uncertainties in the fitted parameters of the spatial-distribution formulae (see Equation~\ref{eq:sigma-binney}) 
are dominated by two parts: 
the statistical uncertainties of the kinematic quantities described in the above,
and the physical fluctuations 
(i.e., deviations from the model owing to astrophysical reasons, 
on small spatial scales, say, with $R \lesssim 1$ kpc and $z \lesssim 0.5$ kpc).
In the analysis of this study (concerning data binning, etc.), 
the two parts are comparable to each other.
And, when used in our Jeans-equations tests (\S\ref{sec:results_JETs}), 
these uncertainties are relatively minor compared with 
the uncertainties in the density profile of tracers (see \S\ref{sec:results_common_tracers}). 
We have checked that our results are not sensitive to binning schemes (including bin sizes and 
the aforementioned thresholds) or fitting methods.
The details of the data binning and analysis are beyond the scope of this study, 
and will be included in a future paper
investigating the Galactic kinematics of Gaia stars.

The tilt angle information (required in the Jeans equations \ref{eq:Jeans_R} and \ref{eq:Jeans_z})
is taken from the measurement by \citet{everall_tilt_2019}
for a sample of disk stars with Gaia DR2 astrometry,
$\alpha = (0.952 \pm 0.007) \arctan(|z|/R)$.

\subsection{Rotation curves and other data} \label{sec:other_data}

\begin{figure*}
	\centering \includegraphics[width=7in]{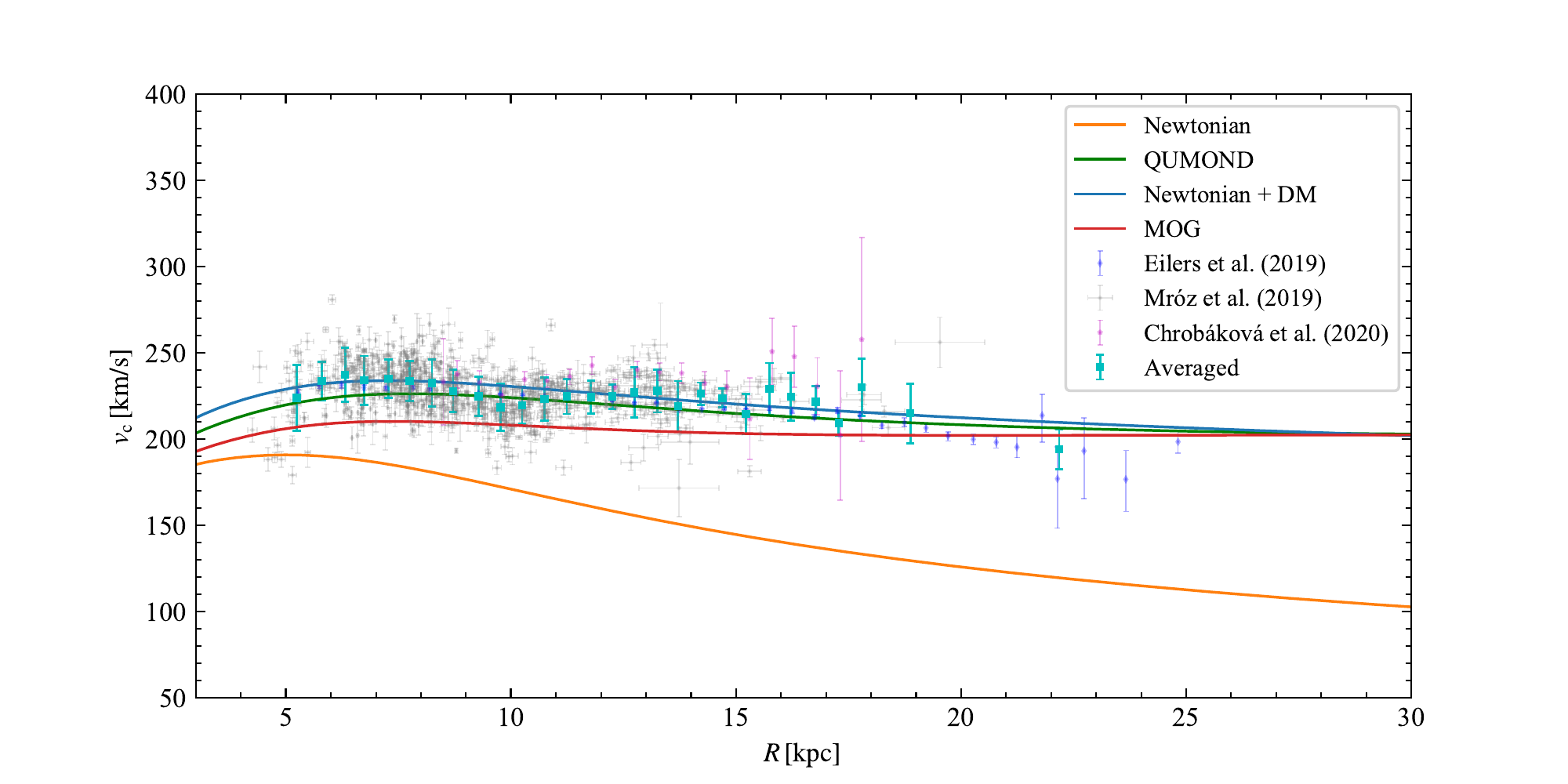}
	\caption{Rotation curves of the Milky Way, with the predicted ones of the gravitational models compared with the observations. 
		The cyan data points with error bars ($\pm 1\sigma$) are our averaged rotation curve 
		over spatial bins with $\Delta R = 0.5$kpc (note that we increase the bin size at a few large-$R$ bins), 
		based on the data of
		\citet[][]{eilers_circular_2019,mroz_rotation_2019,chrobakova_gaia-dr2_2020}. 
		The baryonic mass parameters is from \citet{wangMilkyWayTotal2022}. 
		The orange, blue, green, and red curves represent the Newtonian baryon-only, DM, QUMOND, and MOG models, respectively. 
	    \label{fig:rcsrm-W21}  }
\end{figure*}

The rotation curve data are from giant stars \citep[][]{eilers_circular_2019}, Classical Cepheids \citep[][]{mroz_rotation_2019}, 
and the compilation by \citet{chrobakova_gaia-dr2_2020}. 
They are all consistent with the Galactic constants we use, $R_\odot=8.122~\rm kpc$ and $v_\odot \simeq 229~\rm km\,s^{-1}$.
We note that large scatters exist in the measured circular velocity between different works.
Therefore, we compile the rotation curve by averaging $V_{\rm c}$ over bins of $\Delta R = 0.5~{\rm kpc}$ generally,
and increase the bin size at large $R$ to ensure sufficient S/N (see Figure~\ref{fig:rcsrm-W21}).
The typical (mean) 1-$\sigma$ error of the binned data is 12.1 \kms.  
The size of binning, based on our tests, does not impact our conclusions.

We also used the RAdial Velocity Experiment (RAVE) data \citep[][]{binney_galactic_2014}
to perform the Jeans-equations tests (see below),
and find a good consistence (within $1\sigma$ confidence)
between the results based on the RAVE and Gaia data.

\begin{table}
  \centering
  \caption{Best-fit values of the parameters defined by Equation (\ref{eq:sigma-binney}) required to fit the dependence on $(R, z)$ for $\sigma_R,\,\sigma_\theta,\,\sigma_z,\,V_\theta$, respectively.}
  \label{tab:ourParameters-KinematicFormulae}  
  \begin{tabular}{lcccc}
  \hline \hline
  ~                         & $a_1$    & $a_2$    & $a_3$    & $a_4$ \\ \hline
  $\sigma_R$           & 1.177     & 0.688    & 32.196 & 0.105         \\
  $\sigma_\theta$           & 0.698     & 0.661     & 9.437    & 0.509            \\
  $\sigma_z$      & 0.615     & 0.631      & 34.453  & 0.168          \\
  $V_\theta$           & 6.914       & 0.008        & 2.418    & -0.742           \\ \hline
  \end{tabular}
\end{table}

\section{Results and Discussion} \label{sec:results}

\subsection{Rotation-Curve test}
\label{sec:results_RC} 

We compare the rotation curves predicted by models, $V_{\rm c}(R) = \sqrt{R\,\partial \Phi/\partial R}$, to the observations. 
Figure~\ref{fig:rcsrm-W21} shows the results.
As expected, the Newtonian baryon-only model under-predicts $V_{\rm c}(R)$ evidently, deviating from every binned data points 
by $\gtrsim 3\sigma$ generally.
By adding a DM halo component, the fiducial MW model (see \S\ref{sec:fiducialMWmodel}) 
appears to match the data well, within the $1\sigma$ errors of almost all the bins.
This is also the case for QUMOND.
The MOG model appears broadly consistent with the data, albeit not as good as the fiducial DM model and QUMOND
and systematically smaller than most of the binned data points and the other two gravitational models.

In order to quantify how well the model predictions are consistent with the data, 
we calculate the reduced \chisq\ with a degree of freedom $\rm{d.o.f.} = 27$ regarding to the 28 radial bins. 
Obviously, the Newtonian baryon-only model is rejected by the data with
\QesComm{ $\chi_{\nu, \mathrm{N}}^2 = 51.8 \gg 1$}.  
The fiducial DM model agrees with the data with $\chi_{\nu, \mathrm{DM}}^2 = 0.5$;
QUMOND is broadly consistent with the data with 
QesComm{ $\chi_{\nu, \mathrm{QUMOND}}^2 = 1.5$},
and MOG is also acceptable with $\chi_{\nu, \mathrm{MOG}}^2 = 6.6 \sim \mathcal{O}(1)$, in contrast with the Newtonian baryon-only case.
These $\chisq$ calculations are consistent with the above 
visual impression from Figure~\ref{fig:rcsrm-W21}.

As mentioned in \S\ref{sec:fiducialMWmodel},
We have tested the four gravitational models with other prescriptions of the baryonic mass distributions, 
and with other RC data collected from the literature (see the Appendix), 
and found that all the tests give conclusions similar to the above.

\subsection{Jeans-equations tests}
 \label{sec:results_JETs}

We use Jeans-equations tests to examine how well the gravitational models agree with the data 
outside the Galactic plane. 
According to Equations~\ref{eq:TR} \& \ref{eq:Tz}, 
we calculate $T_R$ and $T_z$ based on three-dimensional kinematic data.
Then we compare them to the respective radial and vertical components of the potential gradients 
predicted by the gravitational models 
(namely, $\partial \Phi / \partial R$ and $\partial \Phi / \partial z$).

$T_R$ and $T_z$, the observed radial and vertical accelerations, 
are derived from the tracer's density profile and kinematic data. 
Their uncertainties ($1\sigma$) are estimated in terms of standard error propagation, as follows:
\begin{equation}
\epsilon_{T_R} = \sqrt{
	\left(\frac{\partial T_R}{\partial \rho}\right)^2\epsilon^2_{\rho}
	+\sum_{i}\left(\frac{\partial T_R}{\partial X_i}\right)^2\epsilon^2_{X_i}
} ~,
\end{equation}
and
\begin{equation}
\epsilon_{T_z} = \sqrt{
	\left(\frac{\partial T_z}{\partial \rho}\right)^2\epsilon^2_{\rho}
	+\sum_{i}\left(\frac{\partial T_z}{\partial X_i}\right)^2\epsilon^2_{X_i}
} ~.
\end{equation}
Here, $\{X_i\}$ are the observed quantities, and $\{\epsilon_{X_i}\}$, their uncertainties. 
We also include a nominal uncertainty of $20\%$ for the tracer's density at each location $(R, z)$.

As for the density profile of the tracer stars, we simply exploit 
the (weighted) whole Galactic disk 
(namely geometrical thin$+$thick disks; see their prescriptions in \S\ref{sec:fiducialMWmodel}), 
\QesComm{
but with appropriate proportion between the two disk components:
\begin{equation}
\rho(R, z) = 0.85\times\rho_{\rm d, thin}(R, z) + 0.15\times\rho_{\rm d, thick} (R, z) ~.
\end{equation}  
The proportional factors (0.85 and 0.15) are the fractions in number of the two disk populations
of the red clump stars
according to their chemical classification
(see \S\ref{sec:our_kinematics_formulae}).
}
But we are not sure if, and how well, the red clump stars follow
the spatial distribution of general stars (cf. \citealt{Piffl_Binney_etal_2014MN445});
also not sure how well the chemically classified thin-disk red clump stars 
are consistent with the dynamically best-fit thin disk of \citet{wangMilkyWayTotal2022}.  
Thus in this study, we also use additional possible density profiles for the tracers, 
 and the results are presented in next subsection.

\begin{figure*}
\hspace*{-0.2in}
	\includegraphics[width=7.5in]{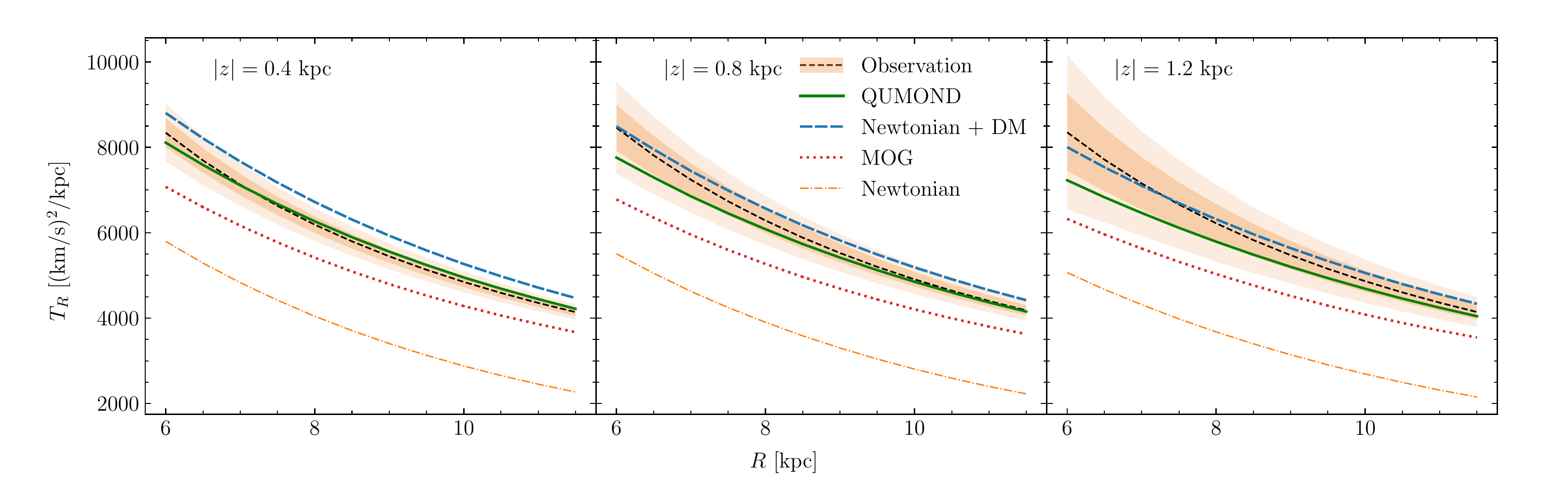} 
	\vspace{-0.2in}
	\caption{Radial Jeans-equation ($T_R$) tests of the gravitational models vs. the data at various $(R,z)$ locations, 
		illustrated as a function of $R$ at different altitudes ($|z|$). 
		In every panel, the dashed black line represents the quantities calculated from the data 
		of the entire Gaia$+$LAMOST sample of red clump stars (\citealt{huang_mapping_2020-1}); 
		Dark and light shades show 68\% and 95\% confidence intervals, respectively; 
		The orange, blue, green, and red curves represent the Newtonian baryon-only, DM, QUMOND, and MOG models, respectively. 
	   \label{fig:TR-W22}  }
\end{figure*} 

\begin{figure*}
	\hspace*{-0.2in}
	\includegraphics[width=7.5in]{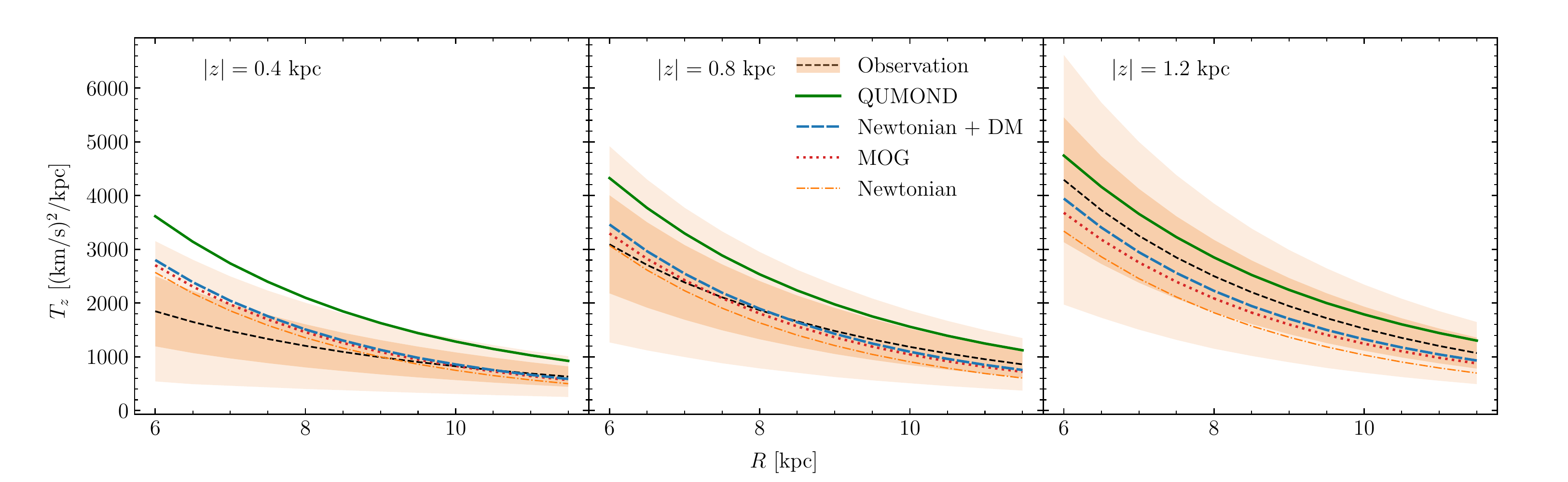} 
	\vspace{-0.2in}
	\caption{As Figure~\ref{fig:TR-W22}, but showing the vertical Jeans-equation ($T_z$) test results. \label{fig:Tz-W22} }
\end{figure*} 

In this subsection, we present the test results for the $(R,z)$ locations 
in the way of illustrating $T_R$ (or $T_z$) 
as a function of $R$, 
at different altitudes (namely distance $|z|$) from the Galactic plane. 
This is because both the Gaia$+$LAMOST data and the RAVE data cover
a limited range in $|z|$.
\QesComm{
The test results based on the RAVE data are consistent with those based on the Gaia$+$LAMOST data.
Because the RAVE-derived $T_R$ and $T_z$ have large errors (see the Figures in the Appendix) 
and might mislead the reader's judgment, we do not plot the RAVE results in the figures of the main text,
but plot them in the Appendix.
}   
Because the calculation of $T_R$ requires $\sigma_\theta$,
the $(R,z)$ space with sufficient data coverage for $T_R$ test 
is $6 < R < 12$ kpc and $-2.5 < z < 2.5$ kpc (see \S\ref{sec:our_kinematics_formulae}).  

Figure~\ref{fig:TR-W22} shows the results of $T_R$ test.
On the observational-data side, $T_R$ monotonically decreases with $R$, 
which just reflects the trend of decreasing magnitude of radial acceleration 
along the radial direction. 
On the model side (the colored lines in the figure), 
generally the radial gradients of the four gravitational models have considerably different magnitudes.
The Newtonian baryon-only model is obviously far below the observed radial acceleration ($T_R$),
for all $(R,z)$ locations.
Likewise, the MOG model is outside at least the 95\% confidence interval of the observed $T_R$,
for all $(R,z)$ locations.
The fiducial DM model basically lies within the 95\% confidence interval of the data
for all the $R$ range at $|z| = 0.8$ kpc (middle panel) and $|z| = 1.2$ kpc (right panel),
except for the case of  $|z| = 0.4$ kpc (left panel) 
where the DM model goes outside the 95\% confidence interval for almost the entire $R$ range.
The QUMOND model behaves best: it lies within the 68\% (1$\sigma$) confidence interval
for almost all $(R,z)$ locations as displayed in the three panels. 

According to Figure~\ref{fig:TR-W22},
one may draw the conclusion that QUMOND fits the data best (within 68\% for almost all spatial locations);
\QesComm{
DM pass the $T_R$ test basically,
at least for all locations with $|z|$ greater than a certain height
(we will see in \S\ref{sec:results_vexing_at_small-z} that in term of $T_R$ test 
the fiducial DM model is outside the 68\% confidence level for all locations at $z \lesssim 0.8$\,kpc);
}
Newtonian baryonic-only model and MOG obviously fail.
Being conservative and for safety, yet we must note that 
the test depends on the tracer's density profile we adopt, and 
that at least DM and QUMOND cannot be discriminated for sure (see next subsection).
\newline  

Independent of the $T_R$ test, we now show the $T_z$ test results in Figure~\ref{fig:Tz-W22}, 
which illustrate the distributions along $R$ direction for the vertical gradients of 
the four potential models at different $|z|$ slices, with respect to the observed vertical accelerations ($T_z$).
While the trend with $R$ is similar to $T_R(R)$,
the magnitude of $T_z$ is in general much smaller than 
$T_R$ of the same locations by at least a factor of 2. 
\QesComm{
	All the four models are broadly consistent with the observations within the 95\% confidence interval.
While the Newtonian baryon-only, fiducial DM and MOG models
lie close to each other, and are all within the 68\% confidence interval for almost all locations,
yet QUMOND lies with the 68\% only at $|z|=1.2$\,kpc (right panel).
To be worse, for the locations at $R <8.5$\,kpc and $|z|=0.4$\,kpc,
QUMOND is outside the 95\% confidence (left panel);
we will see in \S\ref{sec:results_vexing_at_small-z} that in term of $T_z$ test 
QUMOND is outside the 68\% confidence level for almost all locations at $z \lesssim 0.8$\,kpc probably,
and within that confidence for all locations at $z \gtrsim 0.8$\,kpc.  
}

The discriminating power of $T_z$ here is not so strong as $T_R$, as seen from the above test results.
The theoretical reason is that, as mentioned above, in disk galaxies generally
the vertical component of potential gradient 
($\partial \Phi / \partial z$, namely the so-called ``vertical force'' in the literature) 
is much smaller than the radial gradient. 
Thus the differences of vertical field strength between those best-fit gravitational models 
are squeezed together compared with the differences in their radial strength 
(comparing Figures~\ref{fig:TR-W22} and \ref{fig:Tz-W22}).
The observational reason is that
the relative errors (namely the ratios of the aforementioned $\epsilon_{X_i}$ to $X_i$)
of $\sigma_z$ is larger than that of $\sigma_R$ by a factor of $\sim 2$
(cf. Figures~11--14 of \citealt{binney_galactic_2014}),
which are the dominating error terms of the observed vertical and radial accelerations $T_z$ and $T_R$, respectively.
Thus, as displayed in Figures~\ref{fig:TR-W22} and \ref{fig:Tz-W22}, 
the error bars of $T_z$ (and importantly the relative errors) 
are much larger than those of $T_R$ at the same spatial locations.

\subsection{The common results from using different tracers' Density Profiles} 
\label{sec:results_common_tracers}

\begin{figure*}
\hspace*{-0.2in}
	\includegraphics[width=7.5in]{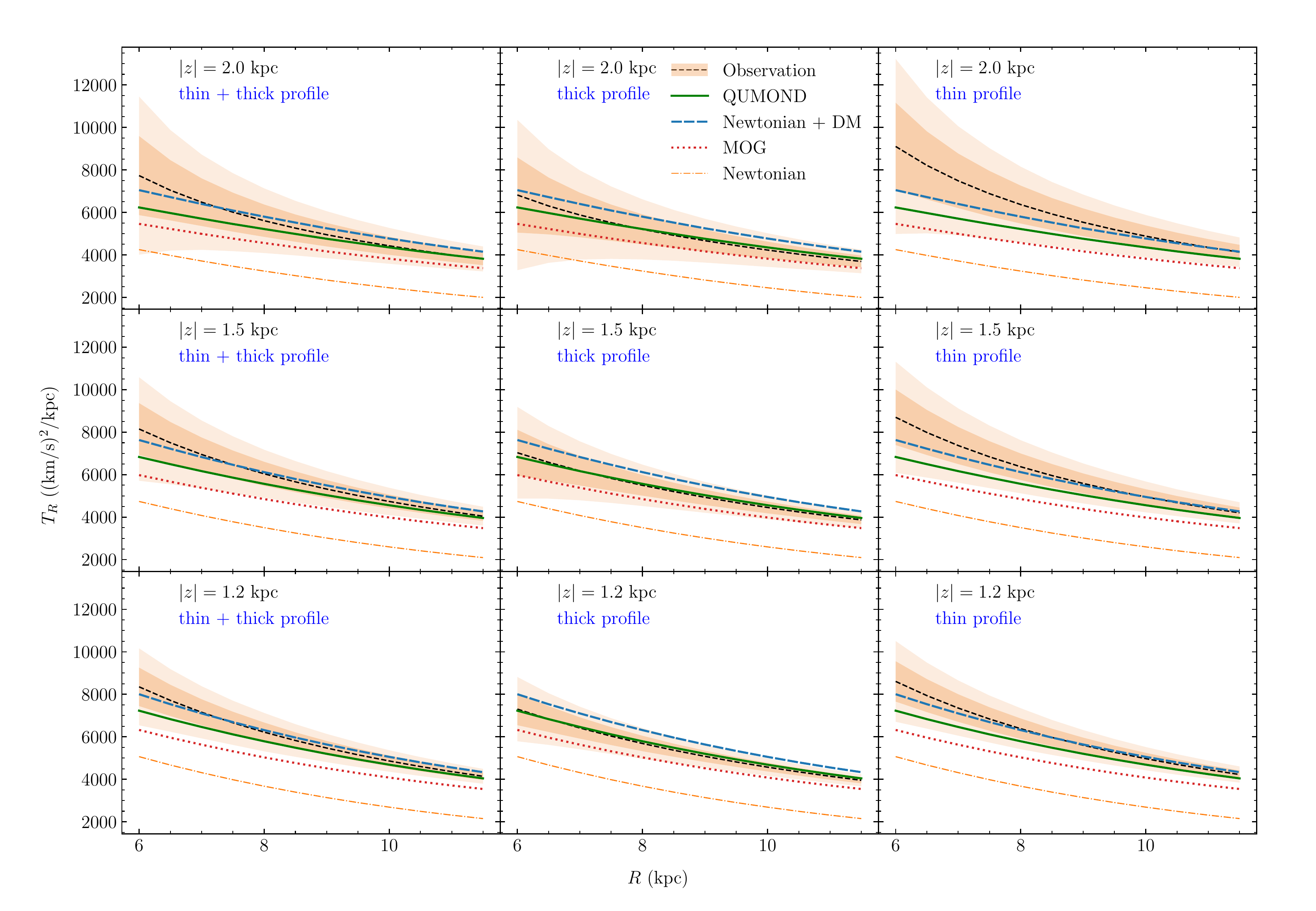} 
	\vspace{-0.2in}
	\caption{$T_R$ test results assuming different tracer's density profiles, 
		based on the entire Gaia$+$LAMOST sample of red clump stars (\citealt{huang_mapping_2020-1}). 
		Three schemes are shown: the total-disk profile (weighted thin$+$thick disks, left panel), 
		thick-disk profile (middle panel), and thin-disk profile (right panel). 
		Denotations are the same as in Figure~\ref{fig:TR-W22}.  \label{fig:TR_tracers-W22} }
\end{figure*} 

\begin{figure*}
\hspace*{-0.2in}
	\includegraphics[width=7.5in]{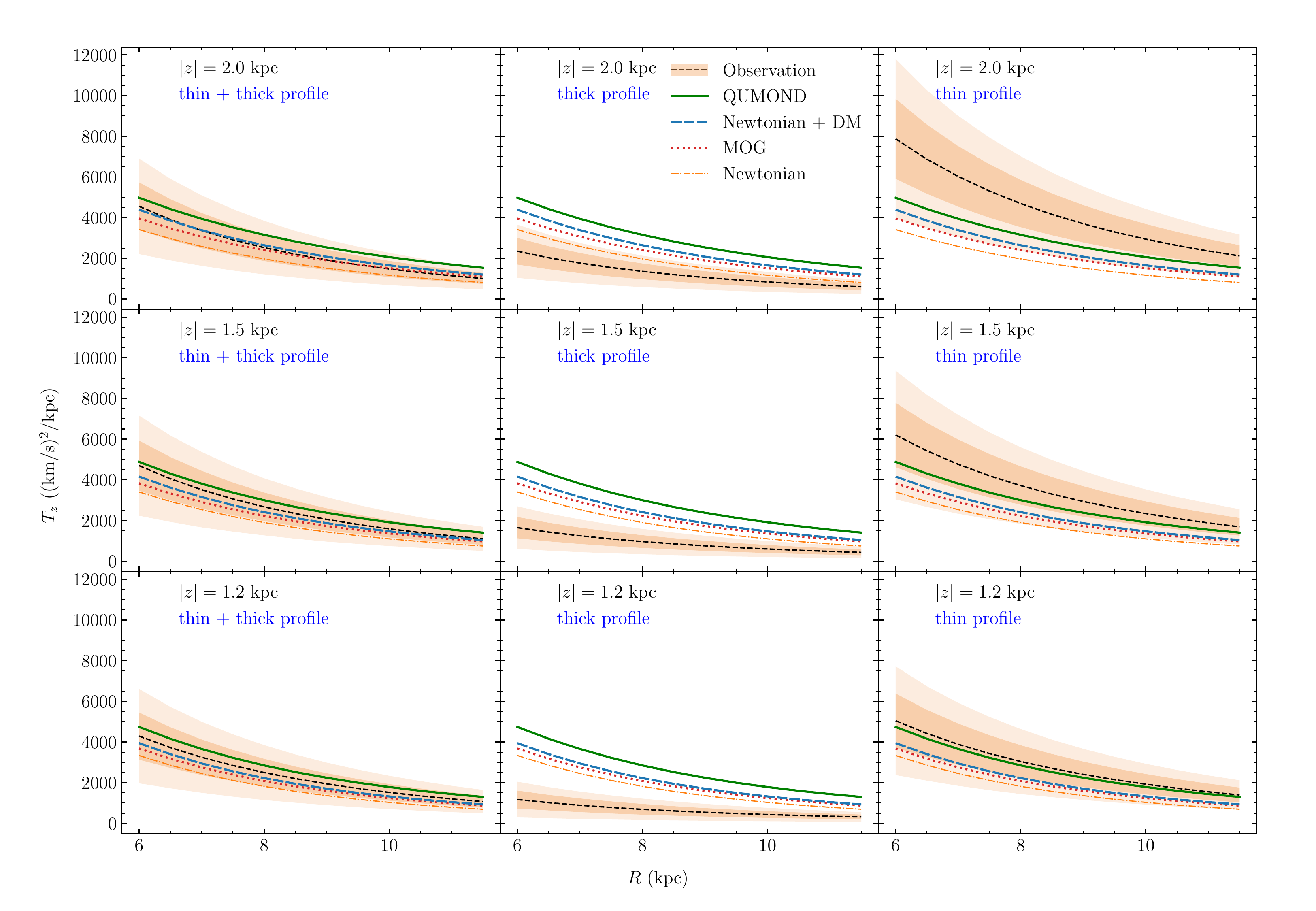} 
	\vspace{-0.2in}
	\caption{As Figure~\ref{fig:TR_tracers-W22}, but for $T_z$ test results. \label{fig:Tz_tracers-W22} }
\end{figure*} 

\begin{figure*} 
\hspace*{-0.2in}
	\includegraphics[width=7.5in]{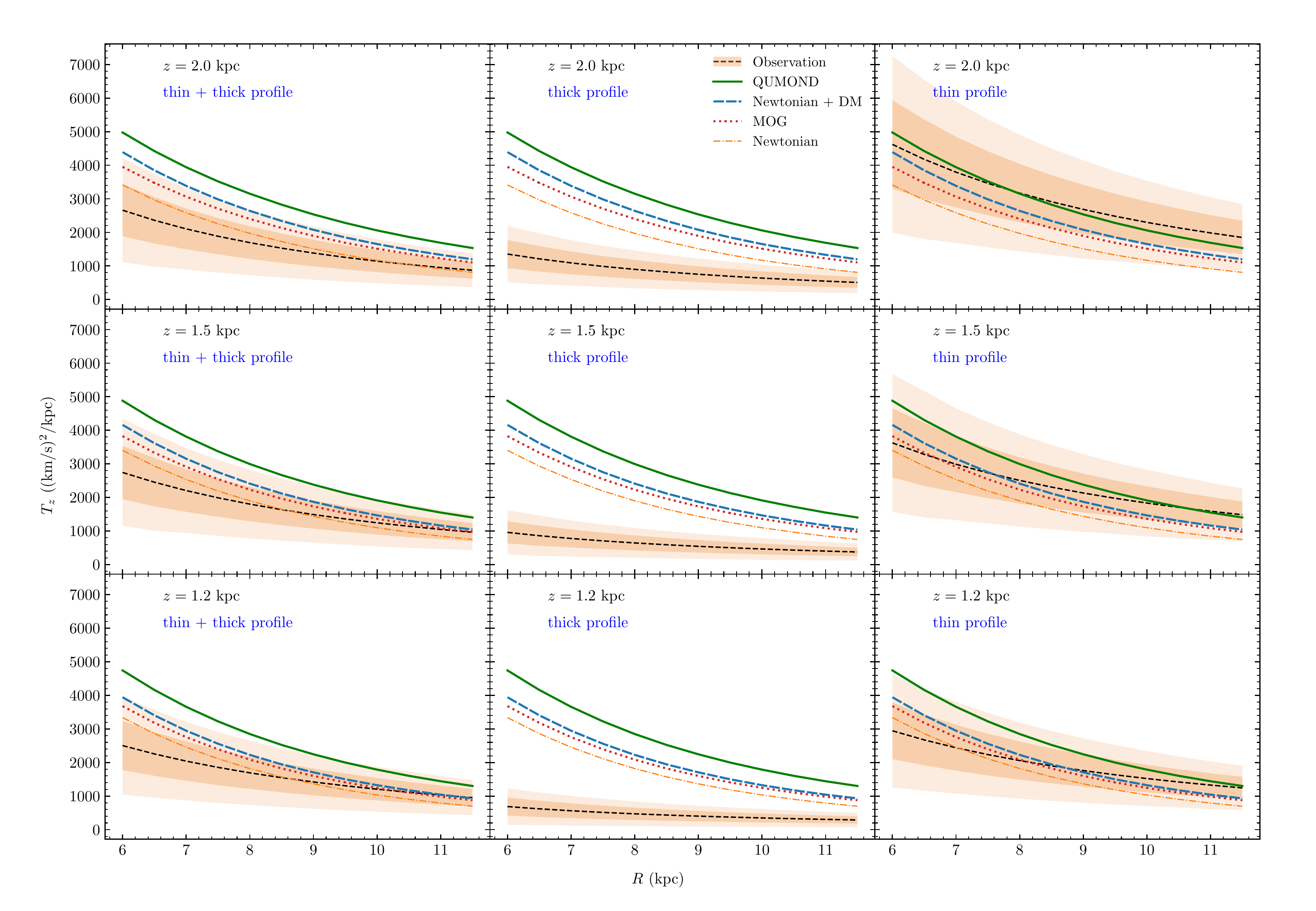} 
	\vspace{-0.2in}
	\caption{As Figure~\ref{fig:Tz_tracers-W22}, but for $T_z$ test results 
		using only the data of the thin-disk red clump stars 
		chemically selected from the \citet{huang_mapping_2020-1} sample.  \label{fig:Tz_RCthin-W22}  }
\end{figure*} 

As stated above,
the major caveat of this work is that 
we lack the knowledge of the shape of the density profile of the tracer population (see $\rho(R,z)$ in \S\ref{sec:jeans}),
and have to represent it by using the profiles of general populations of the disk stars,  
such as the weighted thin$+$thick geometrical disk model of \S\ref{sec:fiducialMWmodel} 
as we adopt in the preceding subsection.
Using different density profiles to represent the tracers' spatial distribution 
may give different values of $T_R$ and $T_z$,
and thus change the relationship between $T_R$ and $\partial \Phi / \partial R$ 
(and between $T_z$ and $\partial \Phi / \partial z$).

In this subsection, we assess the impact of the uncertainty in tracer's density profile
to our Jeans-equations tests, with the following strategy.
We believe that the real shape of the tracer's density profile should be embraced 
by the two main populations of disk stars, namely the geometrical thin and thick disks.
Thus we also use the thin-disk and thick-disk profiles (prescribed in \S\ref{sec:fiducialMWmodel})
to calculate $T_R$ and $T_z$,
and then safely base our conclusions about Jeans-equations tests
on the common results shared by the schemes of using the three kinds of density profiles.

Compared with the above weighted thin$+$thick disk profile scheme,
the thin-disk profile scheme results in larger values of both $T_R$ and $T_z$
(particularly for large-$|z|$ locations),
whereas the thick-disk profile scheme leads to smaller values.
Interestingly, while $T_z$ changes dramatically in the two schemes 
(increased by factors of $1.8 - 3.1$ in the thin-disk scheme, 
and decreased by a factor of $0.4 - 0.8$ in the thick scheme),
$T_R$ changes mildly in the two schemes 
(increased by factors of $1.0 - 1.3$ in the thin-disk scheme, 
and decreased by a factor of $0.9 - 1.0$ in the thick scheme).
That is, while the $T_z$ test is sensitive to the tracer's density profile,
the $T_R$ test is relatively insensitive and thus robust.

The most important $T_R$ and $T_z$ result in common among the three profile schemes 
(excluding the thick-disk profile scheme for $T_z$ test; see the next paragraph),
in a sentence, is the following:
both the fiducial DM model and MOND always lie in 95\% confidence intervals with respect to $T_R$ and $T_z$
for almost all locations with $|z|$ greater than a certain altitude (probably $\gtrsim 0.5$ kpc, see next subsection),
while the MOG model lie farther away from the $T_R$ data at many locations
(let alone the baryon-only Newtonian model).
In addition, there is a second notable point:
On the side of gravitational models, the DM model is always larger than MOND in the radial field strength,
yet always smaller than MOND in the vertical;
what is more, relative to the observed accelerations at low-$|z|$ locations,
the radial field strength of the DM model may even systematically larger than $T_R$ (outside the 95\% confidence)
while the vertical field strength of MOND may even systematically larger than $T_z$ (outside the 95\% confidence),
which will analyzed in detail in next subsection.

The thick-disk profile scheme of the $T_z$ tests yields that 
all the four gravitational models lie beyond the 95\% confidence intervals of the data 
for almost all spatial locations (see Figure~\ref{fig:Tz_tracers-W22}, middle panel).
This fact indicates that the real density profile of the tracers,
i.e., the red clump stars of \citet{huang_mapping_2020-1}, 
is closer to the thin-disk profile than the thick-disk one.
This inference is definitely correct
because, as we recall, the sample of \citet{huang_mapping_2020-1}
is dominated by thin-disk red clump stars (116,000 of 137,000; see \S\ref{sec:our_kinematics_formulae}).
Thus, the $T_z$ tests for the total red clump sample equipped with the thick-disk profile
does not means that this scheme rules out all the four gravitational models,
but means that $T_z$ test is sensitive to tracer's density profile.
This inspires us to consider
the merit of this sensitive dependence in the end of this subsection.

We demonstrate the test results of the two additional schemes (thin-disk profile and thick-disk profile)
in Figure~\ref{fig:TR_tracers-W22} ($T_R$ tests) and Figure~\ref{fig:Tz_tracers-W22} ($T_z$ tests),
together with the weighted thin$+$thick profile scheme as the reference.
To present more new information,
besides the $T_R(R)$ and $T_z(R)$ results of the additional two schemes for $z=1.2$ kpc,  
we plot the results of the three schemes for higher altitudes ($z=1.5$ and 2.0 kpc),
where our data reach 
and the three density profiles for the tracers differ from each other significantly. 
From the figures we can easily see the above-stated features of the test results of the three schemes, 
particularly the most important result in common.

Besides, as already mentioned in \S\ref{sec:our_kinematics_formulae},
we have tried to use only the thin-disk red clump stars chemically selected from 
the \citet{huang_mapping_2020-1} sample 
to perform the $T_R$ and $T_z$ tests.
In this trial, the number of the data points for $T_R$ test 
(i.e., the spatial locations with valid $\sigma_\theta$ and $V_\theta$)
considerably decreases compared with the above analysis,
and thus the power of $T_R$ test is impaired;
the test results of the available data points,
with the thin-disk density profile correspondingly,
are consistent with 
those presented in the right panels of Figure~\ref{fig:TR_tracers-W22}. 
The number of the data points for $T_z$ test decreases not so significantly,
and thus we can perform all the tests, as did in Figure~\ref{fig:Tz_tracers-W22}.
First, of course the $T_z$ tests of the trial case sensitively 
reject the schemes adopting the density profiles of the weighted total disk 
and the thick disk
(see left and middle panels of Figure~\ref{fig:Tz_RCthin-W22}). 
Second, indeed, the trial tests equipped with the thin-disk density profile 
get somehow improved  
than the corresponding ones of the entire-sample case:
MOND and the other three models (the three clustering closely in the $T_z$ plots) 
all together 
lies within the 95\% confidence interval for almost all locations
and even within the 68\% interval for a large fraction of the locations
(please compare the respective right panels of Figures~\ref{fig:Tz_tracers-W22} and \ref{fig:Tz_RCthin-W22}).
Anyway, no matter whether in terms of $T_R$ or $T_z$ tests,
the conclusion remains the same as we conservatively state in the above 
(namely, the result in common). 
\newline  

Concerning the dependence of $T_R$ on tracer's density profile,
we have seen from the above analysis that the dependence is not negligible,
at least for the commonly assumed density profiles in the literature 
(namely the prevailing prescriptions of the Galactic disk components).
Thus we would like to caution that 
if one use $R$-directional Jeans equation to calculate certain quantities 
(e.g., the rotation curves on and off the Galactic plane, \citealt{chrobakova_gaia-dr2_2020}),
the uncertainty caused by tracer's density profile has to be accounted for.

More importantly, concerning the sensitive dependence of $T_z$ on tracer's density profile,
actually there is a potential application.
It is generally difficult to directly derive the density profile of the tracer population (e.g., red clump stars) 
with high completeness (cf. \citealt{Piffl_Binney_etal_2014MN445}).
Instead, if we can constrain the other quantities, i.e., gravitational potential and velocity dispersion, 
then we will be able to place tight constraints on the spatial distribution 
of a specific population of stars (i.e., tracers), 
by taking advantage of the sensitive $T_z$ measure.

In practice, one can even use the two measures in turn as follows. 
First, the $T_z$ measure is employed to pick up plausible models for the tracer's density profile
(based on a grossly correct gravitational model).
Then $T_R$ is used to discriminate various gravitational models with subtle discrepancies.  
The two steps can be iterated to get both the best parameterized 
tracer's density profile and gravitational model.

\begin{figure*}
	\hspace*{-0.2in}
	\includegraphics[width=7.5in]{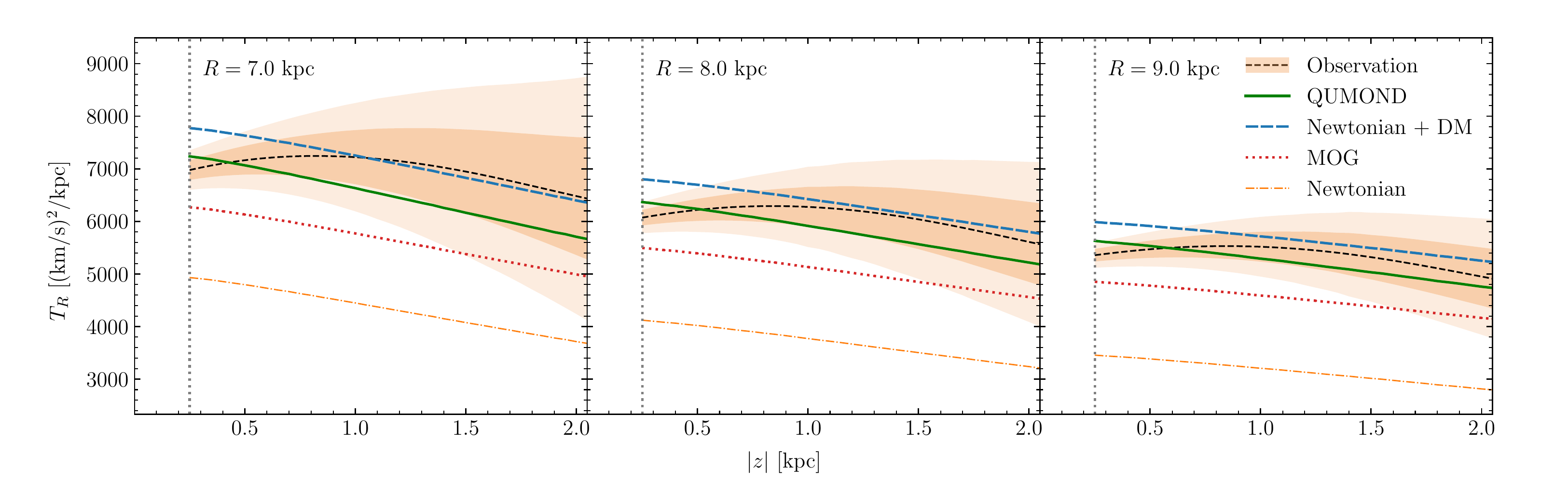} 
	\vspace{-0.2in}
	\caption{Radial Jeans-equation ($T_R$) tests of the gravitational models vs. the data at various $(R,z)$ locations, 
		illustrated as a function of $z$ at three radial positions. 
		In every panel, the dashed black line represents the quantities calculated from the data 
		of the entire Gaia$+$LAMOST sample of red clump stars (\citealt{huang_mapping_2020-1}); 
		Dark and light shades show 68\% and 95\% confidence intervals, respectively; 
		The orange, blue, green, and red curves represent the Newtonian baryon-only, DM, QUMOND, and MOG models, respectively. 
		The vertical grey dotted line denotes the spatial resolution limit in the $z$ direction 
		to the field strengths and observed acceleration ($T_R$).
		\label{fig:TR-z-W22}
	}
\end{figure*} 

\begin{figure*}
	\hspace*{-0.2in}
	\includegraphics[width=7.5in]{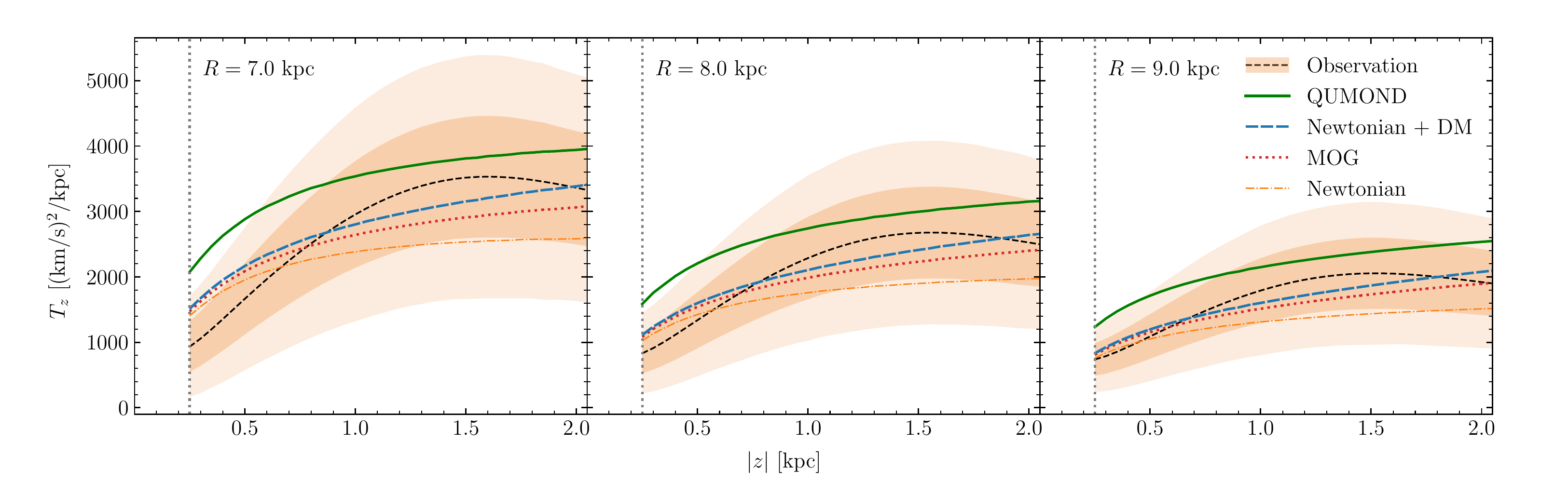} 
	\vspace{-0.2in}
	\caption{As Figure~\ref{fig:TR-z-W22}, but showing the vertical Jeans-equation ($T_z$) test results. \label{fig:Tz-z-W22}
	}
\end{figure*} 

\subsection{Combining radial and vertical dynamics at low altitudes: 
	Vexing for both MOND and spherical DM halos?}  
\label{sec:results_vexing_at_small-z}

From the above two subsections, all the meaningful tests come to a convergent result that
the Newtonian baryon-only model and MOG are rejected, 
and the fiducial DM model and MOND are consistent with the $T_R$ and $T_z$ data generally.%
\footnote{\label{ftn:caveat_baryon_for_MOG-MOND}\QesComm{Concerning MOG and MOND, certainly there is a caveat:
		they are tested by assuming the baryonic matter distribution is the prior one (\S\ref{sec:fiducialMWmodel}) 
		that was best-fitted in the DM paradigm.} 
}
However, there appear systematical trends at low-$|z|$ locations discomforting for both DM and MOND.
For a deeper investigation of the possible low-$|z|$ problem,
in this subsection we plot the $T_R$ and $T_z$ tests as functions of $|z|$, 
at three radial positions $R = 7, 8$ and 9~kpc.
Because the thin-disk stars in our sample are not capable to give $T_R$ tests over large $|z|$ range,
(see also \S\ref{sec:results_common_tracers}),
here we only exploit the test results based on the total sample equipped with the weighted thin$+$thick disk profile.

Regarding the resolution limits in the $z$ direction to $T_R$, $T_z$ and 
the corresponding radial and vertical components of 
field strengths of the four gravitational models 
($\mathbf{g}_{\scriptscriptstyle \rm N}$, $\mathbf{g}_{\scriptscriptstyle \rm DM}$, 
 $\mathbf{g}_{\scriptscriptstyle \rm MOND}$ and $\mathbf{g}_{\scriptscriptstyle \rm MOG}$),
we estimate as follows.
The spatial binning size in $z$ is \mbox{50\,pc} for the kinematic data (see \S\ref{sec:our_kinematics_formulae}),
then according to Nyquist's sampling theorem the resolution limit to the kinematic quantities (e.g., $\sigma_z$)
is twice. 
$T_R$ and $T_z$ involves the first derivative of those kinematic quantities with respect to $z$,
so their spatial resolution limit requires at least two adjacent resolved units, 
i.e., four times the binning size namely \mbox{0.2\,kpc}.
On the gravitational models' side, likewise, the spatial resolution limit to the above field strengths 
is  four times the size of a grid cell, namely \mbox{0.24\,kpc}.
Thus, in the figures we only plot the range from $|z| =$ \mbox{0.25\,kpc} (the resolution) 
to 2\,kpc that our data reliably cover.

In the $T_R$--$z$ plots (Figure~\ref{fig:TR-z-W22}),
the radial field strength of the fiducial DM model lies outside the 95\% confidence interval 
at the locations with $|z| \lesssim 0.5$ kpc,
and does not enter the 68\% confidence until $|z| \gtrsim 0.8$ kpc;
this trend of inconsistency with the $T_R$ data gets somehow worse with $R$ moving outwards.
On the contrary, the radial field strength of MOND always lies in the 68\% confidence interval at every location.

In the $T_z$--$z$ plots (Figure~\ref{fig:Tz-z-W22}),
the vertical field strength of the fiducial DM model always lies in the 68\% confidence interval
of every locations.
On the contrary, the vertical field strength of MOND lies outside the 95\% confidence interval 
at the locations with $|z| \lesssim 0.5$ kpc,
and does not enter the 68\% confidence until $|z| \gtrsim 0.8$ kpc;
this trend of inconsistency with the $T_z$ data gets somehow alleviated with $R$ moving outwards.

In summary, when $|z| \gtrsim 0.8$ kpc both the fiducial DM model and MOND 
lies within the 68\% confidence of $T_R$ and $T_z$ for all locations.
But, at low altitudes (say $|z| \lesssim 0.5$ kpc), there may be problematic:
DM with respect to $T_R$, and MOND with respect to $T_z$.
The exact $|z|$ values have something to do with the tracer population, 
which is subtle to handle as we demonstrated in \S\ref{sec:results_common_tracers};
we defer this issue to future work.
There is a possibility that 
the real Galactic gravitational potential, particularly its inner part, 
is in between the fiducial DM model with a spherical DM halo and the MOND;
that is, in the DM language, the halo may be oblate 
(cf. Figures~\ref{fig:potential} and \ref{fig:pdm} in next subsection).

Note that in Figures~\ref{fig:TR-z-W22} and \ref{fig:Tz-z-W22} 
there are qualitative differences in the shape as a function of $|z|$ 
between the kinematic accelerations (namely $T_R(z)$ and $T_z(z)$\,)
and the field strengths of the four models ($\mathbf{g}_R(z)$ and $\mathbf{g}_z(z)$\,).
The $T_R(z)$ (or $T_z(z)$\,) shapes, in the range shown in the figures, are convex,
while the shapes of the four $\mathbf{g}_{\scriptscriptstyle R}(z)$ (or $\mathbf{g}_z(z)$\,) lines 
look similar and are not so curved.
The reason is that the functions underlying the kinematic and dynamical quantities are different.
The dynamical $\mathbf{g}(z)$ lines are basically determined by 
either the DM halo function (in the case of the DM model) or the baryonic matter distribution (the other three models),
and both the DM halo and baryonic distribution functions decay monotonically farther out (see \S\ref{sec:fiducialMWmodel}).
The kinematic $T_R(z)$ and $T_z(z)$, on the other hand,
are determined by the functions, $\rho(R,z)$ (the tracer's density distribution we adopt)  
and Equation~\ref{eq:sigma-binney} (spatial-distribution function of velocity quantities), and their derivatives; 
the $T_R$ and $T_z$ shapes with respect to $|z|$ are thus complex.
We can imagine, both $T_R(z)$ and $T_z(z)$ would increase rapidly with $|z|$ when $|z|$ larger than a certain value
because of the exponential decay of $\rho(z)$;
this just means that the assumed tracer's density profile, 
likely as well as the extrapolation of the spatial-distribution function of velocity quantities, 
breaks down in that $|z|$ range.
It is right because of the above reason that in Figures~\ref{fig:TR-z-W22} and \ref{fig:Tz-z-W22} 
we only plot the range $|z| \leqslant 2$ kpc (see \S\ref{sec:our_kinematics_formulae}),
and compare the models ($\mathbf{g}$\,) with the data in terms of confidence intervals only.

In the literature it is being hotly debated as to the shape of the Galactic DM halo
is oblate, spherical, or prolate,
with observational evidence both for and against an oblate shape 
of the inner Galactic gravitational potential
(see \citealt{Hattori_Valluri_Vasiliev_2021} and the references therein).
In our above analysis of the possible small-altitude problem,
as the exact $|z|$ range and the degree of DM and MOND deviating from the data
depend somehow on the tracer population and its density profile we use,
thus at this point we leave this problem open. 
\newline  

In the history of MOND research, it is a vexing issue about 
MOND's possible over-prediction of vertical acceleration;
see \S3.1.2 of \citet{Banik_Zhao_2022_review} for a detailed account.
We have noticed that 
\citet{Lisanti_2019_PRD} came to the strong conclusion that 
gravitational models of MOND type failed to simultaneously explain
both the rotational velocity and vertical motion of stars in the solar neighborhood.
In our opinion, there are technical reasons
that explain the tension between their conclusion and our not-so-discriminating one.
There are several problems in their data and modeling method.
The most serious is the key data set they used: 
the observed number density ($n(z)$) and vertical velocity dispersion ($\sigma_z(z)$) 
of three mono-abundance stellar populations at $R = R_\odot$.
The same data set has been thoroughly analyzed by \citet{Budenbender_2015_critic},
which turned out that 
the DM densities estimated by the different stellar populations are inconsistent with each other
(see particularly their Figure~3 and \S3),
owing to a major reason that the data set did not measure 
the cross-dispersion component $\sigma_{Rz}$ of the velocity ellipsoid.
\cite{Hessman_2015_difficulty} also analyzed that data set,
and achieved the same diagnostic as \citet{Budenbender_2015_critic},
along with his other caveats on vertical Jeans-equation modeling;
in fact, as stated in the Introduction, importance of the cross term $\sigma_{Rz}$
has been well proved in past decade. 
By the way, the rotation-curve information \citet{Lisanti_2019_PRD} used 
was limited to a single location, the Solar radius (cf. \citealt{McGaugh_2016_RCgradient}). 
Concerning their modeling method linking $n(z)$ and $\sigma_z(z)$,
which is the completely 1-dimensional Jeans modeling (namely the simplest $K_z$ method),
now it is clear that neither the ``tilt term'' in vertical Jeans equation  
nor the ``rotation-curve term'' in Poisson equation can be neglected 
(see the sixth paragraph of the Introduction and the references therein).

\citet{McGaugh_2016_RCgradient} performed $K_z$ analysis with the ``rotation-curve term'' considered
and yet without accounting for the ``tilt term'' involving $\sigma_{Rz}$ (see his Equation~12), 
based on the $K_z$ (vertical force) data measured by \citet{Bovy_2013_Kz_data}.
The $K_z$ data were derived by complicated action-based distribution function modeling, 
with one assumption being that both $\sigma_R$ and $\sigma_z$ are not dependent on $z$,
i.e., their vertical profiles being constant
(see \S3.1 of \citealt{Bovy_2013_Kz_data}).
A caution mentioned in passing:
the ``rotation-curve term'' in Poisson equation (jargon used in the present paper; also \citealt{Read_2014_JPhG_41})
was called ``tilt term'' in \S4.7 of \citet{McGaugh_2016_RCgradient}. 
\citet{Hessman_2015_difficulty} also critically analyzed the $K_z(z)$ problem of the Bovy \& Rix data set,
along with his comments on the ``accuracy vs. precision'' issue of 
the advanced yet complicated (and thus over-simplified practically) method of distribution function modeling 
(particularly cf. his \S3). 
Recently \citet{Binney_Vasiliev_2022_DF_model} described in detail 
the problems of the (unrealistic) quasi-isothermal distribution function model adopted in \citet{Bovy_2013_Kz_data} 
for Galactic-disk populations.

Besides the above-inspected studies based on the Galactic data,
there are studies based on stellar velocity dispersion ($\sigma_\star$) and other properties 
of galactic-disk stars of external galaxies
(listed in \S3.1.2 of \citet{Banik_Zhao_2022_review}; see also the Introduction).
Just like the \textit{status quo} of those Galactic studies, 
the external-galaxies ones are also inconclusive;
one reason lies in the difficulty of measuring both $\sigma_\star$ and 
requisite other properties (e.g., scale height, or stellar mass-to-light ratio or alike) 
consistently from the same stellar population 
(see, e.g., \citealt{Milgrom_arXiv_critical,Angus_etal_2016_critical,Aniyan_etal_2021}).

\subsection{Exploring the ``extra mass/gravity''}  
\label{sec:results_extra_mass_gravity}

\begin{figure*}
	\centering
	\includegraphics[width=7in]{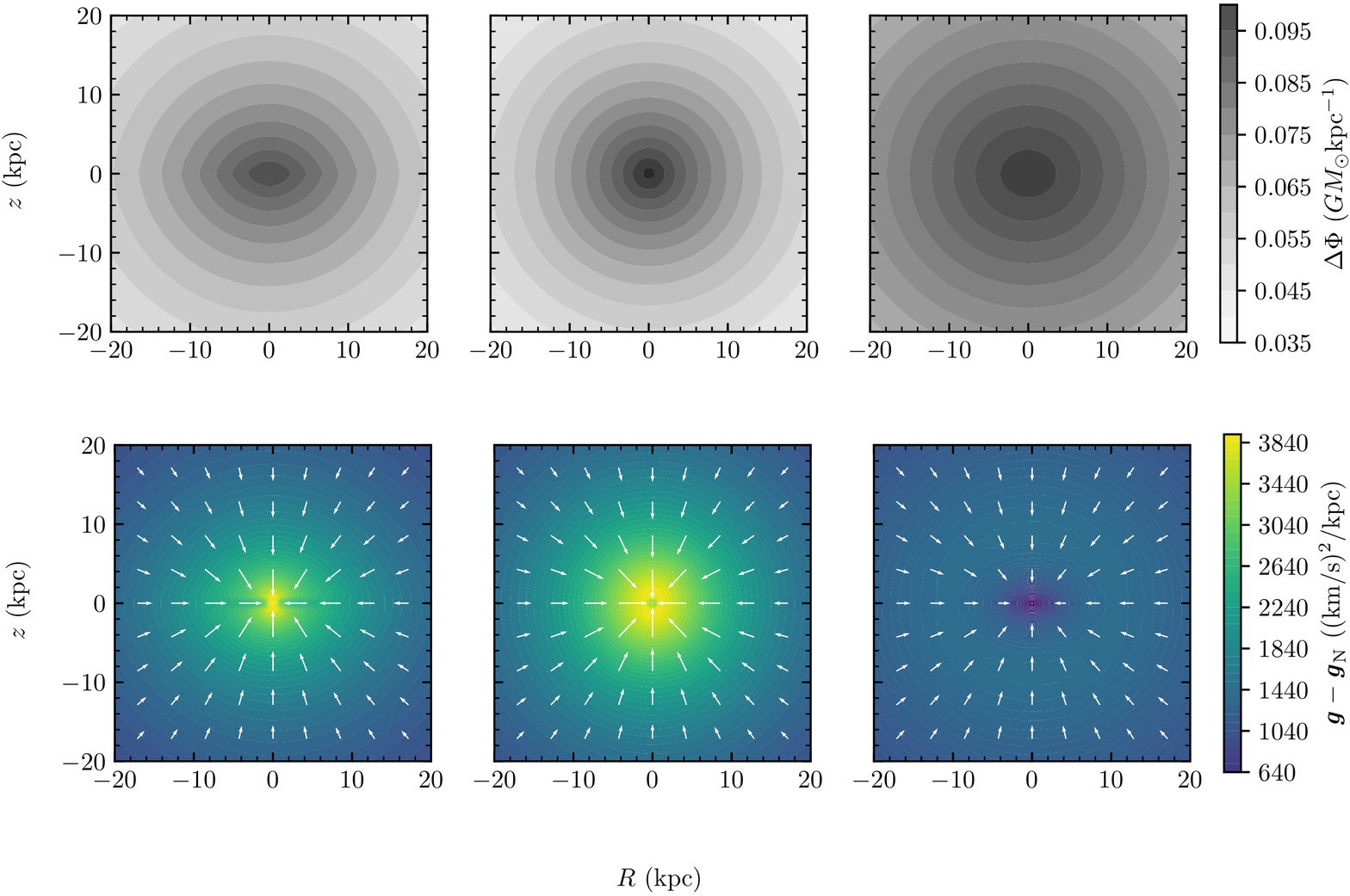} 
	\caption{ \QesComm{\textbf{Top row:}} Gravitational potential difference between the Newtonian baryon-only model and other models 
		($\Delta \Phi = \Phi_{\rm N} - \Phi_{\rm model} $). 
		The left-hand, middle, and right-hand panels are for the QUMOND, DM and MOG cases, respectively.
		Generally the MOG model is not favored by our Jeans-equations tests.
		At present it is yet an open question to what degrees DM and MOND, respectively, 
		represent the real gravitational field of the MW.   
		\QesComm{\textbf{Bottom row:} The corresponding field-strength difference 
			between the Newtonian baryon-only model and other models,  $\mathbf{g}_\mathrm{model} -\gN $. 
			The direction of the vectors is denoted by arrows, and their magnitude is color coded.
		Note that around the center the darker the MOG's extra gravity is (in blue and even purple) 
		the weaker is the field strength. }
	 \label{fig:potential}  }
\end{figure*} 

\begin{figure*}
	\centering
	\includegraphics[width=7in]{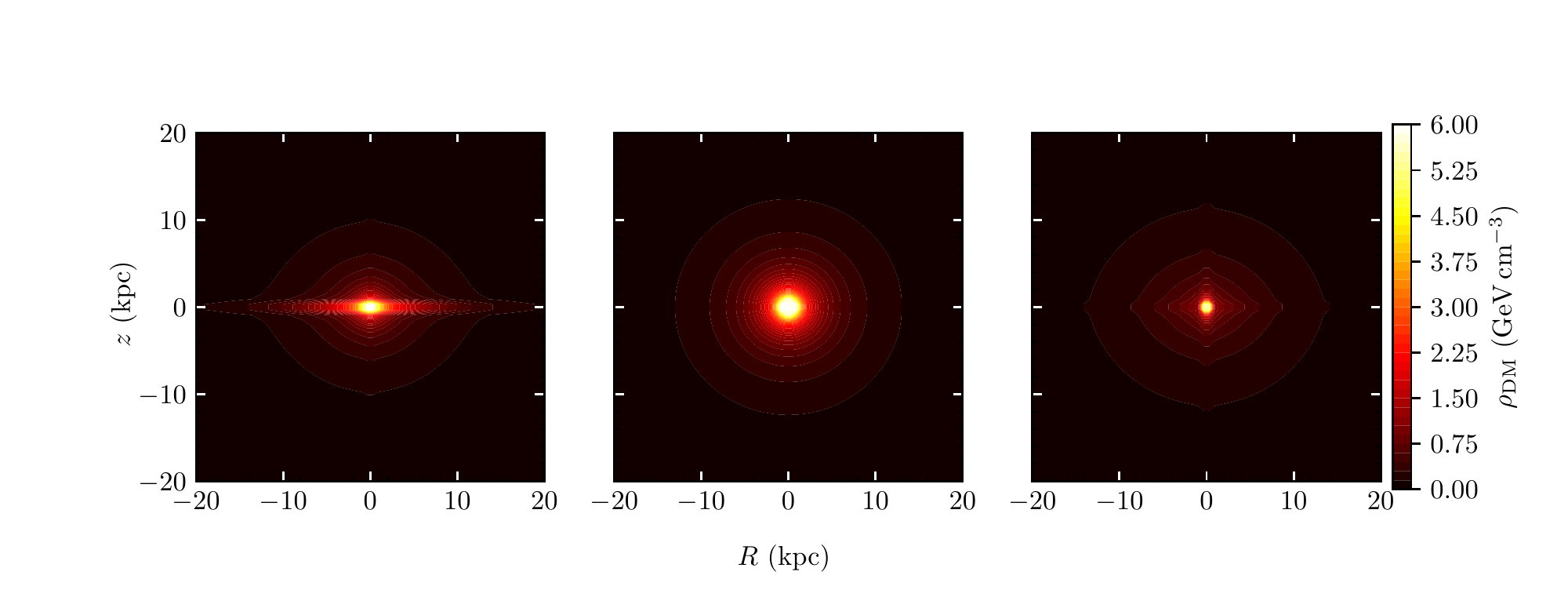}
	\caption{ The ``extra mass'' distribution translated directly from 
		the ``extra potential'' (Figure~\ref{fig:potential}) in terms of the normal Poisson equation.
		{\bf Left:} The density of the effective DM (namely ``phantom dark matter'') predicted by QUMOND. 
		{\bf Middle:} The density of the DM halo in the fiducial mass model.
		{\bf Right:} The ``extra-mass'' density of the MOG case. 
		Caution that the MOG's ``extra mass'' is just a mimic in the DM paradigm, and basically
		useless if not viewed as merely the divergence of the ``extra gravity'' field 
		but interpreted as ``mass'' (see the text for the detail). 
	 \label{fig:pdm}  }
\end{figure*} 

Echoing the early names of the DM problem, 
such as \textit{missing, hidden, excess} or \textit{extra mass} and \textit{excess} or \textit{extra gravity},
with interest we explore the extra mass or extra gravity in excess of the Newtonian baryonic one 
for the DM, QUMOND and MOG models.

We first explore the differences in gravitational potential predicted by the three models
 (denoted as $\Phi_{\rm model}$) 
compared with the Newtonian baryon-only case ($\Phi_{\rm N}$), 
namely $\Delta \Phi = \Phi_{\rm N} - \Phi_{\rm model}$;
\QesComm{also we explore the corresponding gradients of the potential difference,
namely the vector difference in field strength, 
$\mathbf{g}_\mathrm{model} -\gN \equiv \grad (\Phi_{\rm N} - \Phi_{\rm model}).$
Hereafter we call them extra potential and extra gravity, respectively;
yet by definition the two are interchangeable essentially.}
Figure~\ref{fig:potential} plots the distributions of 
the three $\Delta \Phi$ \QesComm{and the corresponding extra gravity} in the meridian plane.
The extra potential of the fiducial model (namely the DM halo; see the middle panel) 
is spherically symmetric as prescribed by the Zhao's profile. 
QUMOND (left panel) gives a comparable extra potential in magnitude to the DM case, 
but the shape of the extra potential is 
\QesComm{fairly flatten in the $z$-direction (i.e., an \textit{oblate} gravitational potential).} 
MOG yields a slightly oblate extra potential (right panel); 
this is clearer in Figure~\ref{fig:pdm}, which can be interpreted 
as the divergence of the ``extra gravity'' field.
In addition, the magnitude of the MOG extra potential
 is $\approx 1.5$ times of the QUMOND or DM one on average.
\QesComm{The magnitude of the extra gravity in MOG is instead 
	fairly smaller than the other two gravitational models (see the bottom row),
which is in fact consistent with the systematic smallness of MOG in the rotation-curve test
(i.e., the gravitational acceleration at $z=0$).}
From \S\ref{sec:results_JETs} and \S\ref{sec:results_common_tracers},
we have seen that
our Jeans-equations tests (mainly the $T_R$) disfavor the MOG model,
yet presently cannot judge for sure 
which one of the fiducial DM model (namely spherical halo) 
and MOND, or some one in between, matches the data to a better degree.

Next, we translate the ``extra potential'' 
(the above $\Delta \Phi$) 
into the effective ``extra mass'' in the Newtonian sense, simply using normal Poisson equation.
In the case of the DM model, this translation is physical and exact; the extra mass is just the DM halo.
We must caution, however, that such a translation is merely mathematical for any modified-gravity models, 
and the concept of ``extra mass'' is even misleading (for the case of MOG; see below).

In the case of QUMOND, interestingly, this translation is meaningful (albeit without any physical content), 
and the ``extra mass'' is the very concept of ``phantom dark matter'' described in \S\ref{sec:qumond}.
This is because 
the QUMOND formulation has a great merit that
its gravitational potential can be naturally decomposed, 
and ascribed in the Newtonian sense to two matter components: the baryonic matter (the real)
and the effective DM (the phantom).
The effective PDM density distribution on the $R-z$ plane 
is shown in Figure~\ref{fig:pdm} (left panel). 
Compared with the density distribution of the DM halo of the fiducial mass model (the middle panel), 
the QUMOND PDM is morphologically closer to 
a traditional (quasi-)spherical DM halo plus a disk-shaped component;
this is consistent with that presented in, e.g., \citet[][]{wu_milky_2008}.

In the case of MOG, just as generic modified-gravity theories 
(e.g., the AQUAL realization of MOND proposed by \citealt{bekensteinDoesMissingMass1984}),
such an ``extra mass'' translation is merely effective; 
i.e., the extra mass distribution (plus the baryonic one) is used in the DM paradigm
to mimic the MOG gravitational potential.
We plot the MOG's ``extra mass'' in the right panel of Figure~\ref{fig:pdm}
just for an intellectual curiosity.
We stress again that our tests are based on MOG's gravitational potential, 
not on the density distribution of the ``extra mass'' described in this subsection.
Certainly, it is correct and useful to view 
the ``extra mass'' (Figure~\ref{fig:pdm}) as the divergence of 
the ``extra gravity'' plotted in the bottom row of Figure~\ref{fig:potential}.

\subsection{On the effective equivalence between MOND and DM} 
\label{sec:effective_equivalence}

After decades of search, DM particles have not been found still \citep{feng_dark_2010}.
Particularly, from the observational standpoint,
the tight correlations between DM and baryonic matter cannot be explained satisfactorily
within the DM framework \citep{bullockSmallScaleChallengesLCDM2017a}.
On the other hand, MOND (or its generally covariant descendants), taken in its present form,
has not been proved to be a mature fundamental theory.
It seems that we still have a long way to go discovering the nature of the ``dark matter problem''.
Just in the above context it is that we are excited by the present study,
tightening the effective equivalence between MOND and DM on circum-galactic and galactic scales,
\QesComm{
or called ``CDM--MOND degeneracy'' (\citealt{Banik_Zhao_2022_review});
to be precise, it is the effective equivalence between the PDM of MOND 
and (possibly oblate) DM halos,
in the sense of acting as gravitational-potential models.} 
A possibility that the effective equivalence is hinting,
\footnote{\label{ftn:effective}
	We must admit that the effectiveness of the equivalence between MOND and DM
	as gravitational-potential models
	is only within the best observational constraints available so far,
	and further tightened by comparing with the MOG case (see Section \ref{sec:results_JETs});
	i.e.,  effective to some degree only.
	Of course, their equivalence is not absolute: 
    as illustrated by Figures~\ref{fig:potential} and \ref{fig:pdm},
the two are different \textit{per} \mbox{\textit{se}. Besides,} 
plausibly they both deserve to be transcended, as discussed in this subsection.} 
is this: A new synthesis may arise, reconciling and transcending both MOND and DM paradigms.
The thinking behind is as follows.
First of all, all the observed correlations between ``DM'' and baryonic matter  
can be explained easily and elegantly
by the simple \citet[][]{milgrom_modification_1983} law (namely the essence of MOND), 
basically without any a free parameter.
This surprising fact suggests the delicate mechanism of the interaction 
between baryons and ``DM'' (particles, or fields, or effective ones)
for the future theory, either in the form of a new gravity 
(say, an effect of quantum gravity, or even a new dynamics/law of nature?),
or in the form of a new ingredient within the established quantum field theory,
or in a third way.
Furthermore, if we take a broader vision, which sees dark energy and DM as two facets of a single origin
as some researchers have pursued \citep[e.g.][]{zhaoDARKFLUIDUNIFIED2010},
then the effective equivalence would point to quantum vacuum, as Milgrom's critical acceleration constant suggests 
(by $2 \pi a_0 \approx c H_0 \approx c^2 \sqrt{\Lambda/3}$\,). 
Finally, we would like to remark that, if there is any minimum value in the above vision,
MOND might be better interpreted as an effect of modified inertia \citep[e.g.,][]{milgromModifiedDynamicsVacuum1999}, 
and even hints at 
nonlocality (nonlocal inertia of \citealt{milgromModifiedDynamicsVacuum1999}, albeit being non-quantumlike for now),  
and reminds us of  
the role of quantum vacuum as ``fluid of virtual particles''. 
Although being exciting, this kind of thinking is speculative so far,
and here we refrain from brain-storming farther.
\footnote{\label{ftn:Hossenfelder}\,
 We want to add a final remark: 
 In all covariant modified-gravity theories so far,
 which are different from the modified-inertia interpretation as Milgrom stressed, 
 one or more additional fields are required;
 those fields have energy and thus are additional sources of gravity, 
 but their stress-energy distribution does not follow that of normal matter 
  (although with other kinds of delicate coupling mechanisms between 
 the additional fields and normal matter; see, e.g., \citealt{Hossenfelder_2017_covariant}).
 Thus, virtually it is interchangeable to call them \textit{additional fields},
 \textit{modification of gravity}, \textit{additional (non-normal) stuff} or directly 
 \textit{non-baryonic DM};
 this is in fact one broader theoretical background inspiring us to think about 
 the effective equivalence between MOND and DM on galactic scales.
 That is, in the direction of modified-gravity interpretation of MOND, 
 the theoretical developments also point to, and have already suggested,
 a transcending synthesis of the two paradigms
 (particularly cf. Section~V of \citealt{Hossenfelder_2017_covariant}).
 After all, from a modern viewpoint of quantum field theory,
 the two paradigms can be conceptually viewed as effective theories 
 for ``collective excitations'' of quantum vacuum (\citealt{Wen_2003_origin}).  
} 

On the other hand, thinking practically, we can exploit the effective equivalence.
As demonstrated in the present study,
for any practical purposes, when researchers want to study the kinematics
on galactic scales, 
they can safely use the QUMOND formula (i.e., the gravitational field of the ``phantom dark matter'')
as an alternative of DM halo models.
This approach will save the researchers from handling various prerequisites and 
fine tuning the cumbersome parameters of DM halos.

\section{Summary} \label{sec:summary}

In terms of the complete form of Jeans equations that admit three integrals of motion,
we perform tests on gravitational models for the Milky Way,
based on the latest three-dimensional (i.e., $R$-, $z$- and $\theta$-directional) 
kinematic data over a large range of $(R,z)$ locations.
Our primary aim is to discriminate between MOND and DM halo models, 
with MOG (as well as Newtonian baryon-only model) as comparison.
The kinematic data we use here are mainly based on the sample of red clump stars 
compiled by \citet{huang_mapping_2020-1},
which are powered by the Gaia DR2 astrometry.

In the Gaia era (from the DR2 onward), previous long-standing problems concerning observational data
(e.g., systematic bias in distance estimation) are gone.
The major factors that affect dynamical modeling of the Milky Way now 
are of astrophysical origin (the complexity of real galaxies), 
e.g., kinematic substructures, 
still rich discrepancies inside a certain tracer population, 
and so on (see \S\ref{sec:our_kinematics_formulae} and \S\ref{sec:results_common_tracers}).
As far as the data we use are concerned,
the typical 1-$\sigma$ error in the rotation-curve data to fit 
is 12 \kms,
and in the velocity-dispersion data 
fitting the spatial-distribution formulae (see below), 0.5~\kms.

Regarding the stellar kinematics 
that we derive based on the data of \citet{huang_mapping_2020-1},
aside from the analytic form proposed by \citet{binney_galactic_2014} for the spatial distributions of $\sigma_R$ and $\sigma_z$,
we find that the spatial distributions of $\sigma_\theta$ and $V_\theta$
also can be well fitted by the same functional expression, 
namely in the form of $\sigma_\theta(R,z)$ and $V_\theta(R,z)$.
We fit the function to the four sets of data, respectively,
and obtain best-fit parameters for the spatial distributions of the four kinematic quantities 
(see Table~\ref{tab:ourParameters-KinematicFormulae}).
We then use the kinematic data calculated in terms of the formulae to perform  
the $T_R$ and $T_z$ tests on every spatial locations.
The advantage is at least two-fold: 
(1) free of the numerical artifacts caused by numerical differentiation given 
the limited spatial resolutions of the observational data,
and more importantly 
(2) reducing the impact of various kinematic substructures in the Galactic disk.

The main results of our comprehensive tests (\S\ref{sec:results_RC}, \S\ref{sec:results_JETs},  
\S\ref{sec:results_common_tracers} and \S\ref{sec:results_vexing_at_small-z})
are summarized as follows:
\begin{itemize}
    \item The Newtonian baryon-only model, as expected,
    is rejected not only by the rotation-curve test (namely dynamics in the Galactic-disk plane),
    but also by the $R$-directional Jeans equation test ($T_R$) for all spatial $(R,z)$ locations.
    \item Concerning the three models with ``extra mass or gravity'' 
    (fiducial DM model with a spherical halo, MOND and MOG),
    rotation-curve data alone (with $z=0$)
    cannot reject any one of them for sure (see Figures~\ref{fig:rcsrm-W21} and \ref{fig:rcsrm_m17}). 
    \item The most important result in common 
    among the Jeans-equation tests with meaningful tracers' density-profile schemes is the following: 
      both the fiducial DM model and MOND always lie in 95\% confidence intervals 
      in terms of both $T_R$ and $T_z$ (the observed radial kinematic accelerations) 
      for almost all locations with $|z|$ greater than a certain altitude ($|z| \gtrsim$ 0.5 kpc probably), 
      while the MOG model lie farther away from the $T_R$ data at many locations 
      (assuming the prior baryonic matter distribution best-fitted in the DM paradigm).
    In particular, both DM and MOND models are equally consistent with the $T_R$ and $T_z$ data within 68\% confidence
    of every locations at $|z| \gtrsim 0.8$ kpc.

\item
At low-$|z|$ locations, there may be problematic trends  
for MOND and the fiducial DM model with a spherical halo, respectively:
the radial field strength of the DM model seems systematically larger than $T_R$ 
while the vertical field strength of MOND seems systematically larger than $T_z$.
To be specific, in the Jeans tests based on the entire red-clump star sample 
equipped with the weighted total disk density profile,
at locations with $0.5 \lesssim |z| \lesssim 0.8$ kpc,
DM is in the 68--95\% confidence while MOND within 68\%  in terms of $T_R$,
and MOND is in the 68--95\% confidence while DM within 68\%  in terms of $T_z$;
at $|z| \lesssim 0.5$ kpc,
DM is outside the 95\% $T_R$ confidence of every locations,
and MOND is outside the 95\% $T_z$ confidence of every locations.
The exact $|z|$ range and the degree of DM and MOND deviating from the data
depend somehow on the tracer population and its density profile,
and thus are uncertain at this point.
There is a possibility that 
the real Galactic gravitational potential, particularly its inner part, 
is in between the fiducial DM model with a spherical DM halo and MOND;
that is, in the DM language, the inner halo may be oblate.
\end{itemize}  
First of all, the above test results consistently point to an observational conclusion:
Even in the condition of current kinematic data 
with the precision and accuracy powered by Gaia DR2
(and the measurement uncertainties are no longer the major concern from now on),
which is able to reject the MOG model (let alone the Newtonian baryon-only model; 
and see the caveat in Footnote~\ref{ftn:caveat_baryon_for_MOG-MOND}),
the MOND model is still not rejected, and behaves as good as the fiducial DM model
through Jeans-equations tests on all spatial locations over 
$5 < R < 12$ kpc and $-2.5 < z < 2.5$ kpc 
(namely the $(R,z)$ space with sufficient data coverage).
This is surprising, because  
(1) there is no free parameter at all in the QUMOND model, 
i.e., without any fitting (let alone pre-fitting), 
and (2) the parameters of the baryonic mass model are actually fine-tuned in the DM context;
on the contrary, 
the fiducial DM model we adopt
was fitted already with all available Galactic kinematic data 
(even the same as part of the rotation-curve data set we use), 
and has been kept improving elaborately for decades.
Secondly, both the fiducial DM model with a spherical halo and MOND
may have their respective vexing facet at low-attitude location (see the forth item above),
which awaits further investigations.

The physical implication of the above test results, what excites us the most,
is the concept that we are tempted to put forward in this paper:
    \textit{the effective equivalence of DM and MOND on circum-galactic and galactic scales}
    (see \S\ref{sec:effective_equivalence} and Footnote~\ref{ftn:effective}).
There may be a value in this concept (as this kind of equivalence is effective 
and hints at both paradigms being effective): 
A new synthesis may arise, reconciling and transcending both MOND and DM.
    On the other hand, from a pragmatic standpoint 
    (\QesComm{the two being equivalent or degenerate gravitational-potential models for now}), 
    we can exploit the effective equivalence in this way:
    when researchers want to study the kinematics
    on galactic scales, 
    they can use the QUMOND formula (i.e., the gravitational field of the ``phantom dark matter'')
    as an alternative of DM halo models.
    This is safe at least on the precision and accuracy level of kinematic data derived from Gaia DR2. 
    This approach will save the researchers from handling various prerequisites and 
    fine tuning the cumbersome parameters of DM halos.

Besides the above astrophysical outputs, 
the present work discovers the instrumental advantages of the two measures, $T_R$ and $T_z$.
The two measures, defined kinematically in terms of the complete form of Jeans equations 
(in axisymmetry at this point; \S\ref{sec:jeans}), represent the observed radial and vertical accelerations 
(fed with kinematic data).  
They both exploit three-dimensional kinematics,
admit three-integral dynamics, and respect three-dimensional Poisson equation.
Thus, first of all, as stated in the Introduction (also in \S\ref{sec:jeans}),
they surpass previous commonly used methods, 
such as rotation-curve test that is essentially a one-dimensional method (namely concerning the $R$-directional dynamics only),
the simple $K_z$ method that is completely one-dimensional also (simplifying a galaxy as $z$-directional slabs),
and most Jeans-equation applications in the literature (assuming dynamics of two integrals of motion only).
More importantly, out of the present work (\S\ref{sec:results_common_tracers}),
we find that $T_R$ test is fairly insensitive to the choice of tracer's density profile
and thus is robust in discriminating gravitational models, 
while the merit of $T_z$ test is instead its sensitivity to tracer's density profile.

Looking forward to the near future, 
we expect to use more (as well as better) kinematic data from, 
e.g., Gaia DR3 astrometry and ongoing large-scale spectroscopic surveys;
importantly, to perform more realistic treatments in galactic modeling
(e.g., handling substructures and refining tracer populations),
and update the Jeans-equations tests of the present study.
Immediately, we would like to make full use of the two measures, $T_R$ and $T_z$,
in the iterative way as described in \S\ref{sec:results_common_tracers}:
first, employ $T_z$ to constrain the model parameters of tracer's density profile
(based on a grossly correct gravitational model);
second, employ $T_R$ to discriminate gravitational models with subtle discrepancies;
then the two steps are iterated to consistently obtain the best realistic
tracer's density profile and best gravitational model.
By doing so, we hope to achieve the final goal:
what a gravitational potential can represent the real Milky Way.

\section*{Acknowledgements}

This work was supported by Natural Science Foundation of China Grants (NSFC 11873083, 11473062). 
\mbox{Y.\,Z.} and \mbox{H.-X.\,M.} were supported by the funding from NSFC for Fostering Talented Students in Basic Sciences 
(through USTC, \mbox{No.\,J1310021}), 
and from CAS for Innovation Training Programs for Undergraduates (hosted by YNAO), 
when they were undergraduate students at USTC. 
Y.H. is supported by National Key R\&D Program of China (No. 2019YFA0405503) and NSFC grants (11903027 \& 11833006).\\
We thank the anonymous referee for the careful reviewing and nice suggestions 
that have been incorporated into and improved the paper, 
and Stacy McGaugh and Moti Milgrom for their helpful comments on the manuscripts from early on;
thank the discussions with James Binney (about tracer's density profile), 
and Xufen Wu (about MOND);
\mbox{X.-B.\,D} thanks his old classmates, Tao TU 
(for his all kinds of knowledge on liquids and solids as well as vacuum, and on foundational statistical mechanics)
and Yifei Chen (for learning the ABC of QFT and general theoretical physics from him)
through the chats in our past half life,
which underlie the theoretical motivation of the present project.

\section*{Data Availability}

The raw data underlying this study are available from the Gaia archive at \url{https://gea.esac.esa.int/archive/}, 
and the LAMOST data releases at \url{http://www.lamost.org/public/?locale=en}\,. 
The distance and 3D velocity data, as well as other information,
of the red clump stars of \citet{huang_mapping_2020-1} we use in this work,
are publicly available by following the link provided by \mbox{Y.\,Huang}: 
\url{https://zenodo.org/record/3875974}\,. 
All the data-analysis code of this work, as well as all data and results at intermediate levels,
are publicly available at \url{https://github.com/ydzhuastro/JeansTest-MW}\,.



\bibliographystyle{mnras}
\input{final_main.bbl}




\appendix
\section{Tests based on the legacy Galactic constants} \label{app:M17}

As we stated in \S\ref{sec:fiducialMWmodel} and \S\ref{sec:results_RC},
We have exploited other parameterizations of the Galactic mass distribution and other kinematic data in the literature,
including those under other sets of the solar position and velocity constants ($R_\odot$ and  $v_\odot$),
and found that our conclusions presented in the main text remain intact.
In the Appendix, we present such an examination: the results based on the mass models and kinematic data 
adopting the legacy Galactic constants $R_\odot=8~{\rm kpc}$ and $v_\odot = 220~\rm km\,s^{-1}$.

In order to use all kinds of rotation curve data in the literature, 
we employ the software {\tt galkin} \citep[][]{patoGalkinNewCompilation2017}.
{\tt galkin} is a powerful tool, that contains the largest compilation of 
rotation curve data of the MW from the literature,
and can bring the data uniformly to a specific Galactic coordinate system the user sets 
(in the Appendeix here, we use the usual Galactocentric cylindrical system with the above legacy Galactic constants).

To be consistent with the legacy Galactic constants,
in this appendix we adopt the Galactic mass model 
that was best fit with the ``Weaker $R_0$ prior'' by \citet{mcmillan_mass_2017},
where the best-fit $R_\odot$ and $v_0$ are consistent with the legacy values within $1\sigma$ uncertainty.
\citet{mcmillan_mass_2017} use the \citet{bissantz_spiral_2002} parametric formula 
for the bulge (with an axisymmetric approximation),
exponential profiles for the stellar disks,
the \citet{dehnen_mass_1998} model for the interstellar medium disks,
and the NFW profile for the DM halo. 
The NFW halo profile is as follows:
\begin{equation} \label{eq:halo-M17} \rho_{\mathrm{h}} =
  \frac{\rho_{0,\mathrm{h}}}{x^\gamma\,(1+x)^{3-\gamma}},
\end{equation}
where $x=r/r_{\mathrm{h}}$, with $r_{\rm h}$ being the scale radius and $\gamma=1$.
The parameterizations of the other profiles are detailed in \S\ref{sec:fiducialMWmodel}.
The specific parameters are listed in Table \ref{tab:model-M17}.
In fact, all main-stream Galactic mass models that prevailed in the literature
have no substantial difference from each other; this is at least true for the goal of our present study.

\begin{table}
  \centering
  \caption{Parameters of the ``Weaker $R_0$ prior'' mass model of \citet{mcmillan_mass_2017}.}
  \begin{tabular}{lcccc}
  \hline \hline
  disk                         & Thin    & Thick    & \HI    & $\mathrm{H}_2$ \\ \hline
  $\Sigma_0 [\!\msun \kpc^{-2}]$ & 9.52e8  & 1.20e8   & 5.31e7 & 2.18e9         \\
  $R_{\mathrm d} [\!\kpc]$       & 2.40     & 3.47     & 7.0    & 1.5            \\
  $z_{\rm d} [\!\kpc]$           & 0.3     & 0.9      & 0.085  & 0.045          \\
  $R_{\rm m} [\!\kpc]$           & -       & -        & 4.0    & 12.0           \\ \hline \hline
  Spheroid                     & Halo    & Bulge    &        &                \\ \hline
  $\rho_0 [\!\msun \kpc^{-3}]$   & 6.98e6  & 1.02e11 &        &                \\
  $r_{\rm h} [\!\kpc]$                 & 21.21 & -        &        &                \\ \hline
  \end{tabular}
  \label{tab:model-M17}
  \end{table}

\begin{figure*}
	\centering \includegraphics[width=7in]{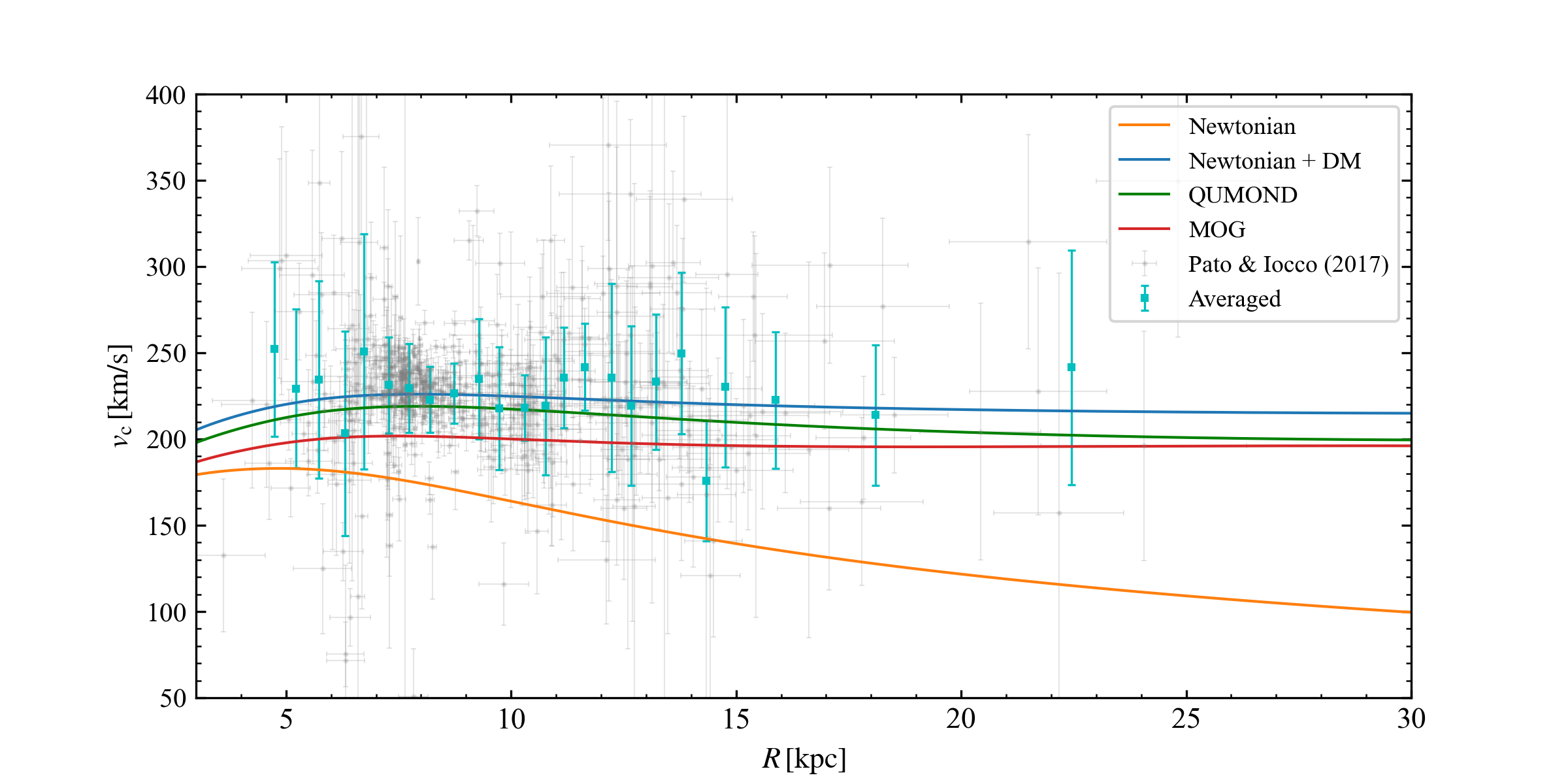}
	\caption{As Figure~\ref{fig:rcsrm-W21}, but for the mass model and observational data based on $R_\odot=8~{\rm kpc}$ and $v_0 = 220~\rm km\,s^{-1}$. 
		The cyan data points with $\pm1\sigma$ error bars are our averaged rotation curve 
		over spatial bins with $\Delta R = 0.5$ kpc (we increase the bin size at large $R$), 
		based on the data compiled by {\tt galkin} \citep[][]{patoGalkinNewCompilation2017}. 
		The baryonic parameters are from
		the ``Weaker $R_0$ prior'' mass model of \citet{mcmillan_mass_2017}. \label{fig:rcsrm_m17}  }
\end{figure*}

Figure~\ref{fig:rcsrm_m17} shows the rotation curve for the ``Weaker $R_0$ prior'' model.
We also calculate the reduced \chisq\ with a degree of freedom 
$\rm{d.o.f.} = 23$ regarding to 24 radial bins of the data. 
The DM, QUMOND, and MOG models have reduced $\chi_{\nu, \mathrm{DM}}^2 = 0.2$,
$\chi_{\nu, \mathrm{QUMOND}}^2 = 0.2$ and $\chi_{\nu, \mathrm{MOG}}^2 = 0.5$.
This is consistent with the visual inspection from Figure~\ref{fig:rcsrm_m17}; i.e., 
all the three gravitational models match most of the binned data points within their $\pm 1\sigma$ measurement uncertainties.
In contrast, not surprisingly at all, 
the Newtonian baryon-only model is not favored with $\chi_{\nu, \mathrm{N}}^2 = 3.4$,
larger than the other three models by an order of magnitude. 
We note that the $\chi_{\nu}^2$ values here 
are smaller than those in Section \ref{sec:results_RC} 
because the measurement uncertainties of the binned data points are large;
this is reasonable considering that the {\tt galkin} compilation is heterogeneous.

\begin{figure*}
    \centering
    \includegraphics[width=6.5in]{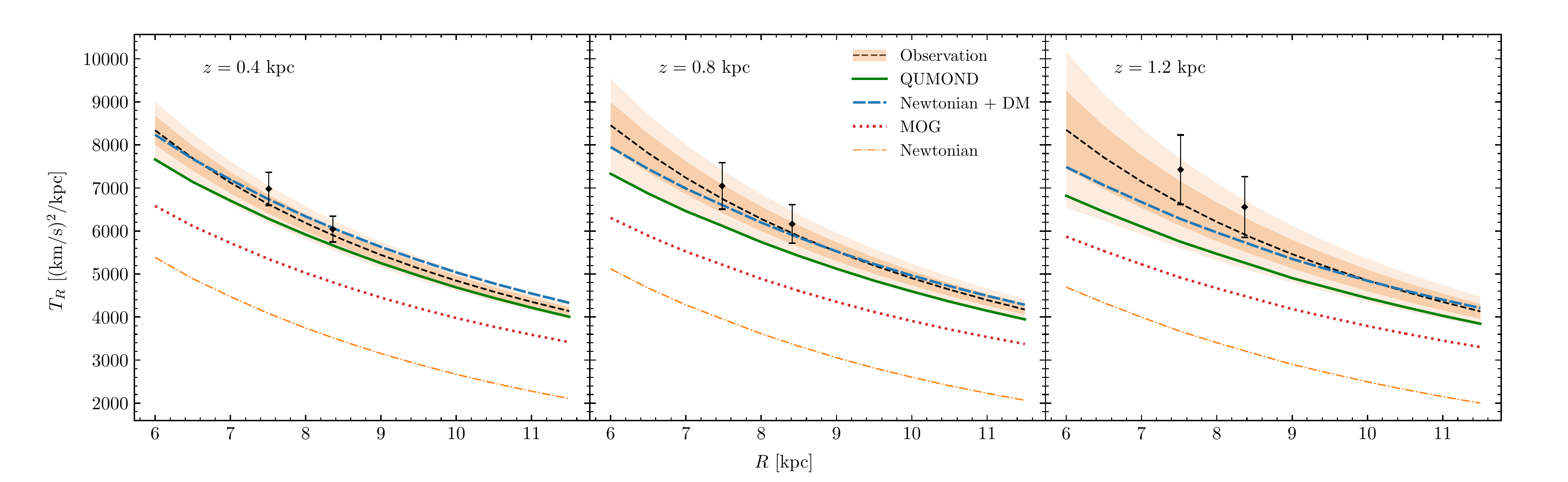}  
    \caption{As Figure~\ref{fig:TR-W22}, but showing the $T_R$ test results for the ``Weaker $R_0$ prior'' mass model of \citet{mcmillan_mass_2017}.
    \QesComm{Also plotted is the results based on the RAVE sample (black dots with $\pm1\sigma$ error bars).}
    \label{fig:TR-M17}   }
\end{figure*}

\begin{figure*}
    \centering
    \includegraphics[width=6.5in]{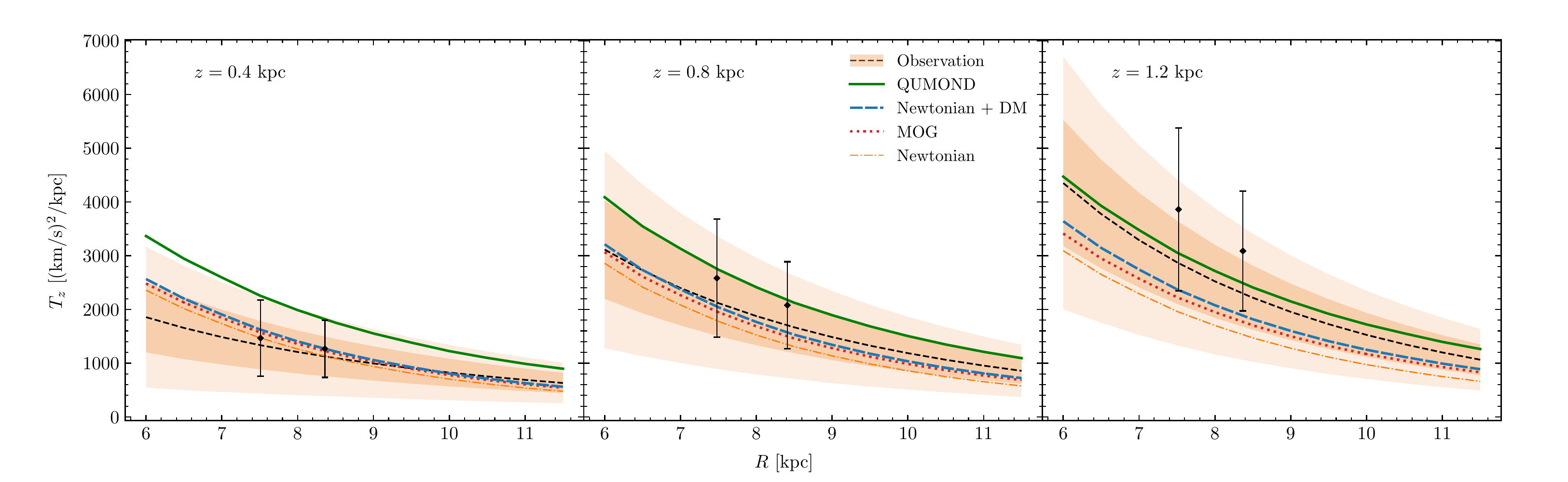} 
    \caption{As Figure~\ref{fig:TR-M17}, but showing the $T_z$ test results. \label{fig:Tz-M17}  }
\end{figure*}

We also present the Jeans-equations tests based on the ``Weaker $R_0$ prior'' mass model of \citet{mcmillan_mass_2017}
and the three-dimensional velocity data as described in Section \ref{sec:data}. 
The results are presented in 
Figures~\ref{fig:TR-M17} and \ref{fig:Tz-M17}. 
As mentioned in \S\ref{sec:results_JETs},
we also show the RAVE-based results whose $T_R$ and $T_z$ have large errors.
It is evident that these new $T_R$ and $T_z$ tests are also consistent 
with our results in \S\ref{sec:results} based on the W22 mass model.

\bsp	
\label{lastpage}
\end{document}